\documentclass[12pt,preprint]{aastex}
\usepackage{amsmath}

\usepackage{color}
\usepackage{rotating}

\def\simlt{\lower.5ex\hbox{$\; \buildrel < \over \sim \;$}}
\def\simgt{\lower.5ex\hbox{$\; \buildrel > \over \sim \;$}}
\newcommand{\itb}{{\it b}} 
\newcommand{\ee}[1]{\mbox{${} \times 10^{#1}$}}
\def\kms{\mbox{km s$^{-1}$}}
\def\dotdeg{\mbox{$.\!^\circ$}}
\def\msol{{M$_\odot$}}

\def\vexp{v_{\rm exp}}
\def\vlsr{v_{\rm LSR}}

\def\schi{{\sc Hi}\ }
\def\schii{{\sc Hii}\ }
\newcommand{\Ht}{{\rm H$_2$}}

\def\g54{{G$54.4-0.3$}}
\newcommand{\igalfa}{{I-GALFA}}

\newcommand{\vgps}{{VGPS}}
\newcommand{\cgps}{{CGPS}}
\newcommand{\bonn}{{Effelsberg}}  

\newcommand{\iras}{{\em IRAS}}
\newcommand{\fermi}{{\em Fermi}}
\newcommand{\agile}{{\em AGILE}}
\newcommand{\rosat}{{\em ROSAT}}
\def\tsf{t_{\rm sf}}

\def\esn{E_{\rm SN}}

\def\vdet{$v_{\rm det,min}$}                  

\shorttitle{}
\shortauthors{G. Park et al.}%

\begin{document}
\title{\schi Shells and Supershells in the \igalfa\ \schi 21-cm Line Survey: I.
Fast-Expanding \schi Shells Associated with Supernova Remnants}

\author{G. Park\altaffilmark{1}, 
        B.-C. Koo\altaffilmark{1,13},
        S. J. Gibson\altaffilmark{2}, 
        J.-h. Kang\altaffilmark{3,4,5}, 
        D. C. Lane\altaffilmark{3,6}, 
	K. A. Douglas\altaffilmark{3,7},
        J. E. G. Peek\altaffilmark{8,9},
        E. J. Korpela\altaffilmark{10},
        C. E. Heiles\altaffilmark{11},
        J. H. Newton\altaffilmark{2,12}
        }

\altaffiltext{1}{Department of Physics and Astronomy,
                 Seoul National University,
                 1 Gwanak-ro, Gwanak-gu, Seoul 151-742,
                 Republic of Korea}
\altaffiltext{2}{Department of Physics and Astronomy,
                 Western Kentucky University, Bowling Green,
                 KY 42101, USA}
\altaffiltext{3}{Arecibo Observatory, HC 3 Box 53995, Arecibo, PR 00612, USA}
\altaffiltext{4}{Yonsei University Observatory, Yonsei University,
                 50 Yonsei-ro, Seodaemun-gu, Seoul 120-749,
                 Republic of Korea}
\altaffiltext{5}{Korea Astronomy and Space Science Institute,
                 776 Daedeokdae-ro, Yuseong-gu, Daejeon 305-348,
                 Republic of Korea}
\altaffiltext{6}{Department of Electrical and Computer Engineering,
                 San Diego State University, San Diego, CA 92182, USA}
\altaffiltext{7} {University of Calgary/Dominion Radio Astrophysical Observatory, P.O. Box 248, Penticton, BC V2A 6J9, Canada}
\altaffiltext{8} {Department of Astronomy, Columbia University, New York, NY 10027, USA}
\altaffiltext{9} {Hubble Fellow}
\altaffiltext{10} {Space Sciences Laboratory, University of California, Berkeley, CA 94720, USA}
\altaffiltext{11} {Radio Astronomy Lab, UC Berkeley, 601 Campbell Hall, Berkeley, CA 94720, USA}
\altaffiltext{12}{Department of Physics and Astronomy, McMaster University, Hamilton, Ontario L8S 4M1, Canada}
\altaffiltext{13}{Corresponding author; koo@astro.snu.ac.kr}

\begin{abstract}

We search for fast-expanding \schi shells associated with Galactic 
supernova remnants (SNRs) in the longitude range $\ell\approx$ 32\arcdeg\ to 
77\arcdeg\ using 21-cm line data from the Inner-Galaxy Arecibo L-band Feed 
Array (\igalfa) \schi survey. Among the 39 known
Galactic SNRs in this region, we find such \schi shells 
in four SNRs: W44, \g54, W51C, and CTB~80. 
All four were previously identified in low-resolution 
surveys, and three of those (excluding \g54) were previously studied
with the Arecibo telescope. 
A remarkable new result, however, is the detection of \schi emission 
at both very high 
positive and negative velocities in W44 from the 
receding and approaching parts of the \schi expanding shell, 
respectively. {\em This is the first detection of both 
sides of an expanding shell associated with an SNR 
in \schi 21-cm emission.} The 
high-resolution \igalfa\ survey data also reveal
a prominent expanding \schi shell with high circular symmetry 
associated with \g54. 
We explore the physical characteristics 
of four SNRs and 
discuss what differentiates them from other 
SNRs in the survey area. 
We conclude that these four SNRs are likely 
the remnants of core-collapse supernovae
interacting with a relatively dense ($\simgt 1$~cm$^{-3}$) 
ambient medium, and we discuss the visibility of SNRs in the \schi 21-cm line.
 
\end{abstract}

\section{Introduction}
\label{sec:intro}

The interstellar medium (ISM) is pervaded by small and large
expanding neutral atomic shells 
\citep[e.g.,][]{heiles79,heiles84,mcclure02,ehlerova05,mcclure12}. 
These \schi shells are the interstellar
material swept up by supersonic shock
waves produced by mechanical energy sources, including
\schii regions, stellar winds, supernova (SN) explosions, 
and infalling high-velocity clouds.
The dominant and most violent 
sources are supernovae (SNe), which dump huge amounts of kinetic energy
into the Galactic ISM every 20--70~yrs.
But it is not clear how this kinetic energy is conveyed to 
the diffuse ISM, because this depends on the types and 
the physical environments of SNe.
Most supernovae are core-collapse SNe (CCSNe) that have 
massive ($\ge 8$~\msol) progenitors, and 
most CCSNe are produced in clusters \citep[e.g.,][]{higdon05}. 
Most SN explosions, therefore, are  
correlated in both space and time, with such groupings frequently producing 
supershells and superbubbles with radii of more than a few hundred parsecs.
Only Type Ia SNe and a small fraction of CCSNe are likely to occur in 
isolation.  
 Single CCSNe probably explode inside a wind bubble created by 
their progenitor stars during the main-sequence phase. 
For stars of spectral type later than B0, 
this bubble size is small ($\simlt 1$~pc), and the 
SNRs can interact with dense molecular clouds in their early evolution
\citep{chevalier1999}. 
Type Ia's, on the other hand, probably explode in either 
warm diffuse environments of density $n \sim 0.1~$cm$^{−3}$ 
or in hot, rarefied gas with $n \sim 10^{−3}$~cm$^{−3}$.
The amount and characteristics of the 
kinetic energy imparted to the ISM by SNe, therefore, 
should be diverse, and their role in shaping the 
kinematics of the 
atomic phase of the ISM is not clear.
Consequently, \schi observations of shells and supershells are useful
not only to understand the nature and origin of
individual structures but also, 
with reasonably large statistical samples,
to explore the overall effects of SNe on the ISM.

There have been a number of systematic searches for \schi shells
associated with individual Galactic SNRs.
\citet[hereafter KH91]{koo91} carried out a survey of Galactic SNRs in
\schi 21-cm line using the Hat-Creek 25~m telescope (FWHM$=36'$).
They observed
103 northern Galactic SNRs and detected high-velocity (HV) gas
toward 15 SNRs including three SNRs known prior to the survey.
\cite{koo04b} searched for similar \schi features toward 97 southern SNRs
using the Parkes data from the 
Southern Galactic Plane Survey \citep[FWHM$=16'$;][]{mcclure01}
and identified another 10 SNRs.
Since the SNRs are usually less than 1\arcdeg\ in diameter, high-resolution
observations are essential to confirm the association
of HV \schi features with the radio continuum SNR. 
Such confirmations have been
made in several cases, e.g.,
CTB~80 \citep{koo90}, W44 \citep{koo95a}, W51C \citep{koo97a},
and IC~443 \citep{gio79, bra86, lee08}.
There have also been studies of almost stationary \schi shells or \schi bubbles
associated with SNRs, but it is generally difficult to derive the parameters of
such low-velocity \schi shells due to \schi background confusion (e.g.,
\citealt{kothes05}; \citealt{cazzolato05}; 
see also references in \citealt{koo04b}).

Recently, the 7-beam Arecibo L-band Feed Array (ALFA) receiver on the Arecibo
305m telescope has enabled Galactic \schi\ surveys of unprecedented breadth and
sensitivity with a fully-sampled 4\arcmin\ beam \citep{peek10,peek11}.
The Inner Galaxy ALFA (\igalfa) survey \citep{koo10,gibson12}
covers the portion of the first Galactic quadrant visible to Arecibo, an area
of more than 1650 square degrees at longitudes of 32\arcdeg\ to 77\arcdeg\ in
the Galactic plane and extending to 10\arcdeg\ or more off the plane. \igalfa\ 
uses 0.184~\kms\ velocity channels over an LSR velocity range of $\sim -700$ to
$+700$~\kms.  Its brightness temperature RMS noise is 0.2~K in single empty
channels.  This combination of
high sensitivity, high spatial and spectral resolution, and large area and
velocity coverage are well suited for a systematic study of \schi\
shells and supershells in the diffuse interstellar medium. 
In this first
paper, we search for fast-expanding \schi\ shells associated with {\em known} 
SNRs and consider the implications of our results on their nature.  A
forthcoming paper will discuss known \schi\ shells and supershells as well as
newly identified shells in the \igalfa\ survey data.
It is worth noting that the VLA Galactic Plane Survey \citep[\vgps;][]{stil06}
and Canadian Galactic Plane Survey \citep[\cgps;][]{taylor03} 
cover most of the SNRs in the first quadrant 
in HI at higher spatial resolution ($\sim$~1\arcmin--2\arcmin).  
Their velocity coverage and sensitivity ($\pm$~100--150~\kms\ 
and 1--3~K per 0.8~\kms\ channel) 
are not ideal to study faint, fast-expanding HI shells, 
but they can be useful to study shell fine structure 
at relatively low velocities.

In Section~2, we explain how we identify 
fast-expanding \schi shells associated with SNRs and 
present the resulting list. In Section~3, 
we summarize the results on four SNRs that have  
associated fast-expanding \schi shells.  The two SNRs with new results, 
W44 and \g54, are discussed in some detail.
In Section~4, we 
explore the physical characteristics 
of the four SNRs and compare their properties to the other 35 SNRs
in the survey area. 
Section~5 summarizes the paper. 

\section{Identification of SNR \schi shells}
\label{sec:identify}

There are 275 Galactic SNRs in Green's catalog \citep{green09a,green09b}.
Among them, 39 SNRs are included in the \igalfa\ area (Table~\ref{tab:all}).
The SNR name, size, and type parameters 
in Table~\ref{tab:all} are from \citet{green09a}. 
(See the table note for more parameter details.) 
Note that 
the 21-cm spectra of sources with high continuum brightness are noisy,
so faint high-velocity emission could not be seen in those objects.
In Table~\ref{tab:all}, the ranks determined by KH91 are also listed. 
KH91 observed each SNR with the Hat Creek 85-foot telescope (FWHM=36$'$) 
at 9 points in a cross
pattern centered on its catalog position and searched for 
SNRs with broad ($\simeq 10$~\kms) excess emission over the background.
They divided SNRs into four ranks: 0, 1, 2, and 3, where rank~3 has the highest 
probability 
for an associated fast-expanding 
\schi shell.  In rank~1, the central excess emission is
brighter than the four outermost positions; 
in rank~2, the central excess emission is 
brighter than all outer positions; 
in rank~3, there is also 
excess emission at the highest positive or negative velocities;
and in rank~0, none of these criteria are met. 
Among 39 sources in Table~\ref{tab:all}, 26 were studied by KH91, 
who classified 9 of these as rank~3 SNRs.
KH91 mapped the excess HV emission in these 9 SNRs 
and concluded that the \schi emission is not physically associated in 4 cases
(marked as `(3)' in Table~\ref{tab:all}).
Four SNRs are in rank~2. 
The other 13 (ranks~1 and 0) did not show any significant excess \schi 
emission.

In order to identify fast-expanding shells associated with SNRs, 
we examined whether enhanced emission is present at the SNR position 
at high positive/negative velocities in several ways.
We first inspected average, background-subtracted  
\schi spectra toward individual SNRs as KH91 did.
The average spectrum was obtained from  
a circular area with 1.1 times the SNR radius, while 
the background spectrum was obtained 
from an annular ring of 3\arcmin\ thickness surrounding the SNR. 
The radial interval between the SNR circle and the background ring
was either 9\arcmin\ or 0.5 times the SNR radius, whichever was smaller.
If another bright radio continuum source was located
near the SNR, we left 
an appropriate space between the SNR circle and the background ring.
We confirmed that all SNRs classified as rank~3 by KH91 show
excess emission at high positive/negative velocities in their 
background-subtracted spectra,
but we could not find additional SNRs with such features.

The above approach could have missed associated \schi 
emission limited to small areas. We therefore 
inspected individual channel maps as well as $(\ell,v)$ and $(b,v)$ maps 
to search for HV \schi emission features spatially correlated with the SNRs 
in radio continuum maps.
These are mostly from the \vgps\ or 
\cgps\ 21-cm continuum data with $\sim$~1\arcmin\ resolution.
For SNRs outside these two survey areas, we used 
either the lower-resolution ($\sim4\farcm3$) 
\bonn\ 11-cm continuum data \citep{reich90} or referred to previous works.
Surprisingly, this detailed inspection yielded no additional detections.
Instead, we confirmed the conclusion of KH91 that the HV emission in 
four rank~3 SNRs extended beyond the spatial extent of SNRs, 
so that it was probably 
not associated with the SNRs (marked as `(3)' in Table~\ref{tab:all}). 
In one of them (G$40.5-0.5$), 
we found that the HV \schi emission is part of a much larger and prominent 
expanding shell. This result will be discussed in a forthcoming paper.

We are therefore left with the five rank~3 SNRs of KH91, but we suspect
the HV \schi emission in W50 (G$39.7-2.0$) is probably not associated with 
the SNR. W50 is a large, elongated shell-like SNR with SS~433 at the center,
and KH91 reported the presence of very weak, extended 
\schi emission at high positive velocities
along the northern edge of the SNR where 
the continuum emission is enhanced \citep[cf.][]{lockman07}.
The high-resolution \igalfa\ data confirm the presence of 
weak, filamentary \schi emission along the SNR edge 
that appears connected to other \schi structures 
well outside the SNR. 
We therefore regard the spatial correlation between the \schi emission and 
the SNR as a coincidence.
Our results are summarized in 
the last column of Table~\ref{tab:all}, where we comment only on 
the HV emission features. We do see some low-velocity 
\schi features that may be associated with SNRs, but 
many of these are confusing, and the scope of the present paper is limited to
{\em fast-expanding} \schi shells in SNRs.

\section {Supernova Remnants with Fast-Expanding \schi Shells}
\label{sec:indiv}

\begin{figure}[ht]
\epsscale{1}
\plotone{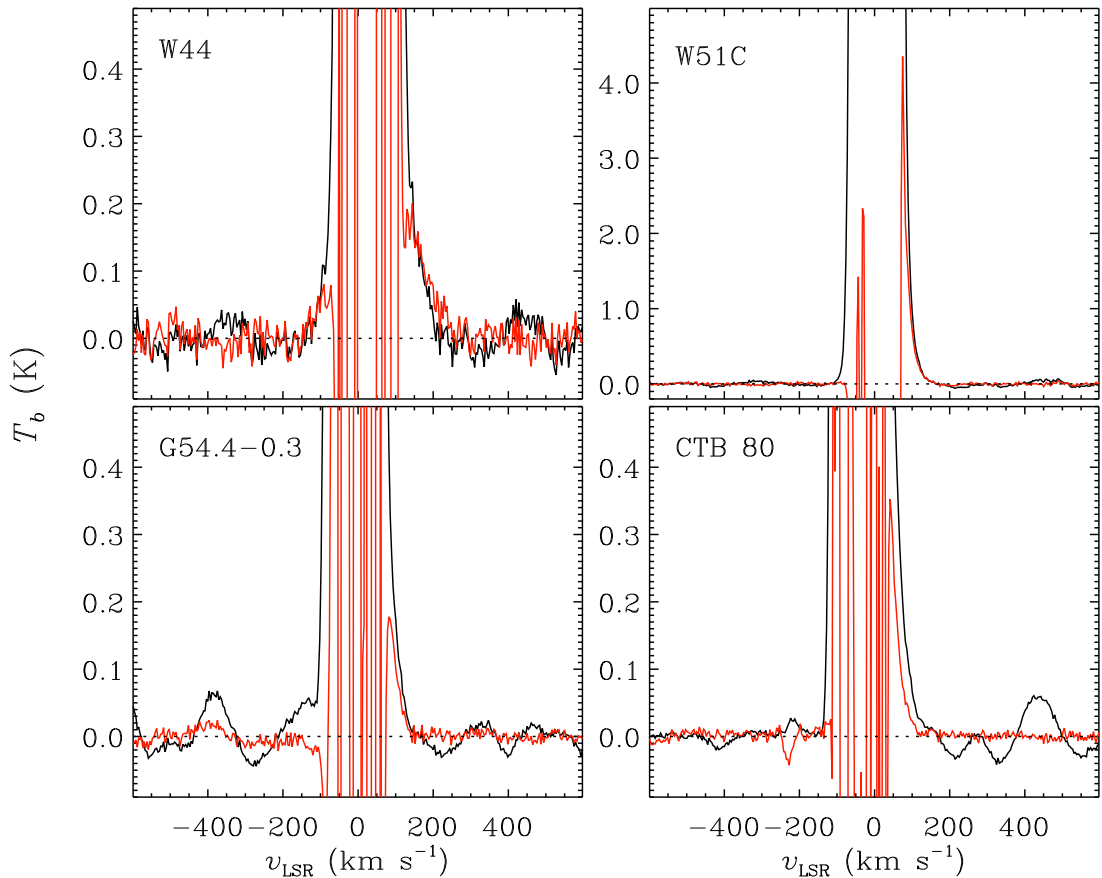}
\caption{
Average \schi 21-cm line profiles of SNRs
that have associated high-velocity \schi emission.
Average profiles toward the SNRs are in black,
while background-subtracted profiles are in red.
(See \S\ref{sec:identify} for details of the background subtraction.)
Variations in \schi 21-cm brightnesses between the source and background 
directions cause the wild fluctuations in the background-subtracted profiles
at low velocities ($|\vlsr| \simlt 100\kms$), but these have no effect on our
analysis.
}
\label{fig:vprof}
\end{figure}

\begin{figure}[ht]
\centering
\plottwo{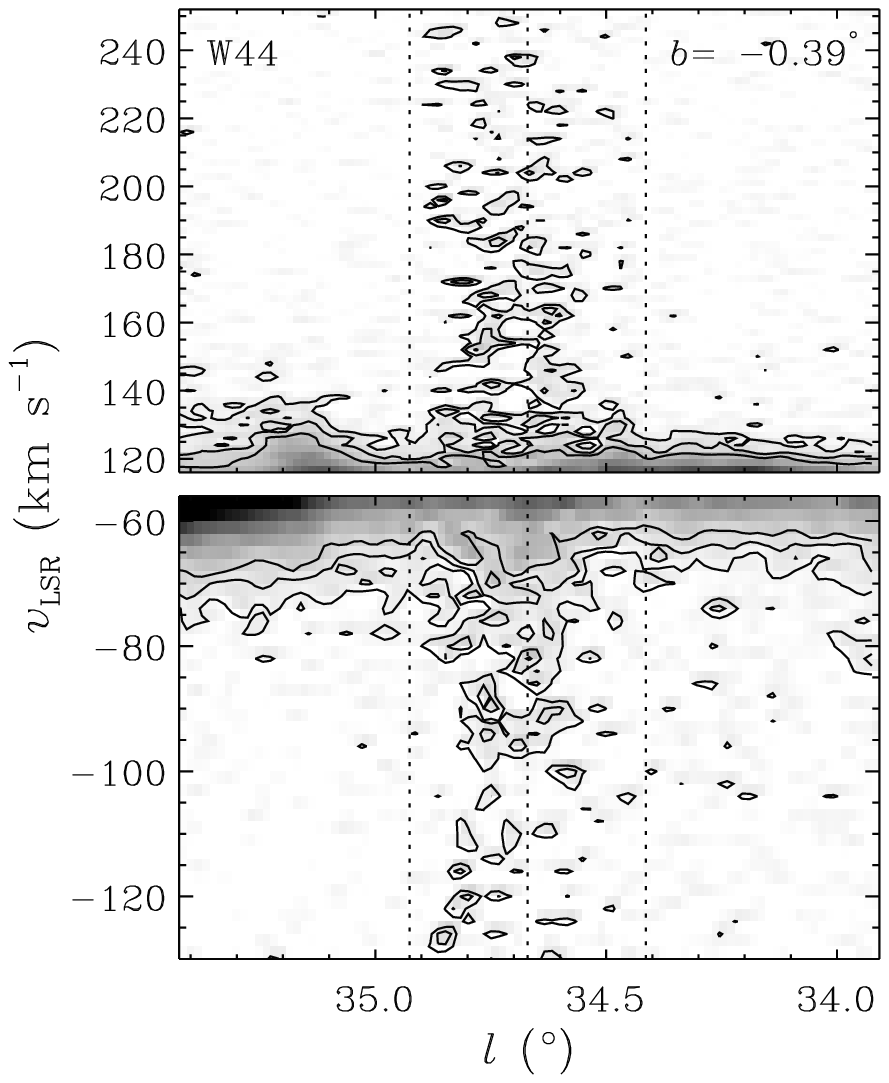}{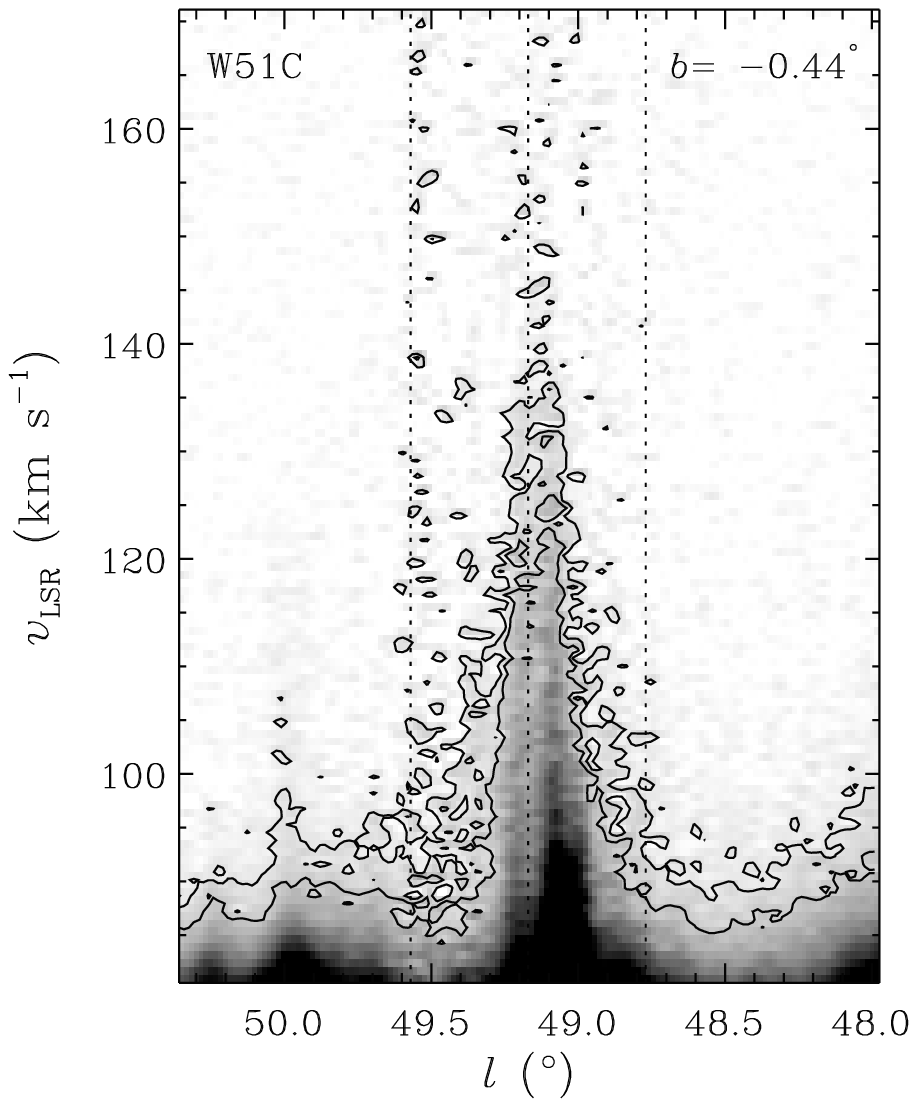}
\plottwo{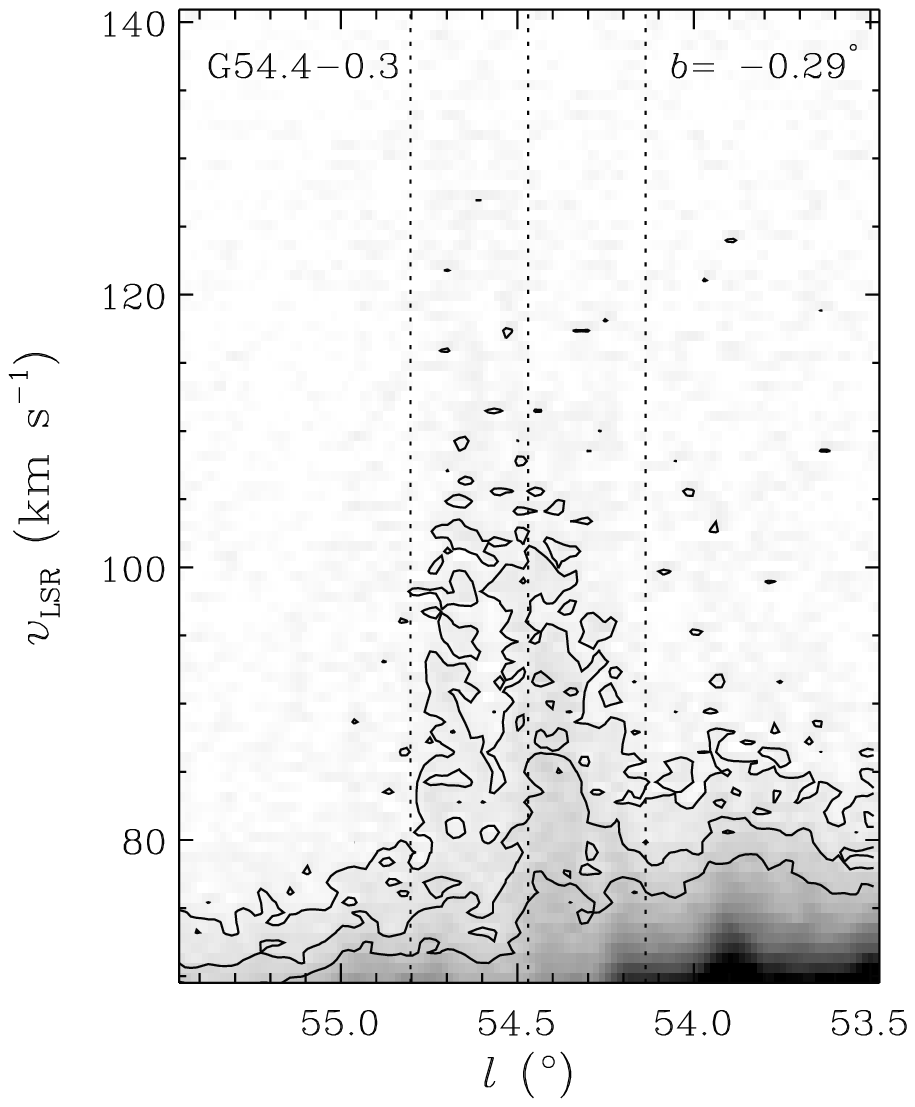}{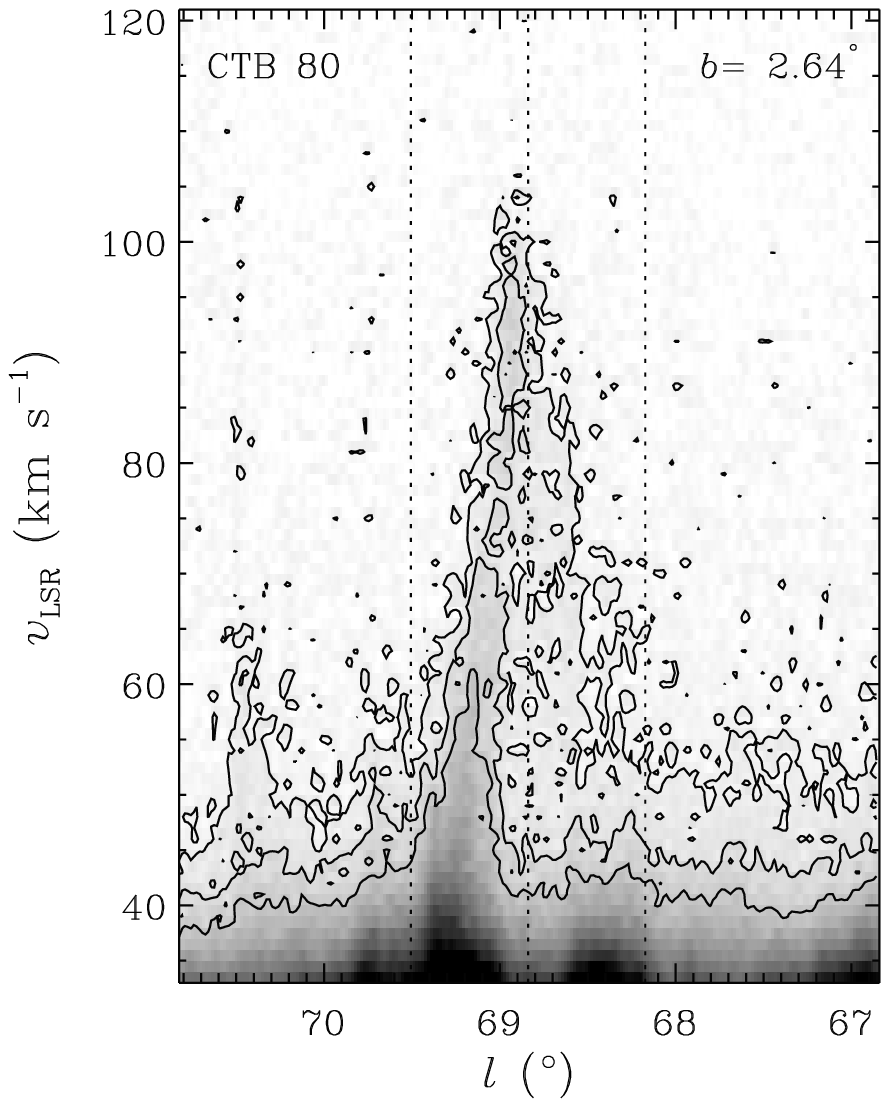}
\caption{
\igalfa\ ($\ell,v$) maps of \schi emission of four SNRs.
Each longitude-velocity map is at the latitude given in its upper-right corner.
The 
central longitude of the SNR is marked by the middle dotted line in each panel.
The other two dotted lines indicate the boundary of the SNR. 
The gray scale varies from 0 to 5~K (white to black).
Contours are 0.5 and 1.0 K for W51C and 0.3, 0.6, and 1.0 K for the other SNRs.
}
\label{fig:lvmap}
\end{figure}

\begin{figure}[ht]
\epsscale{1.0}
\centering
\plottwo{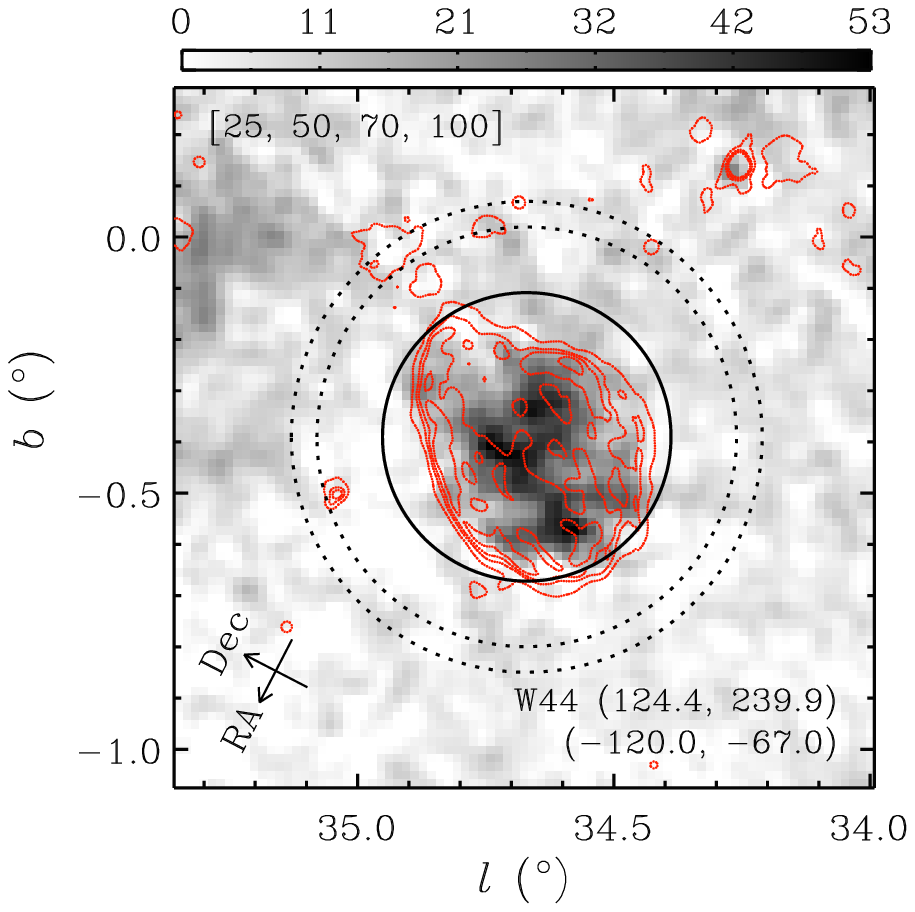}{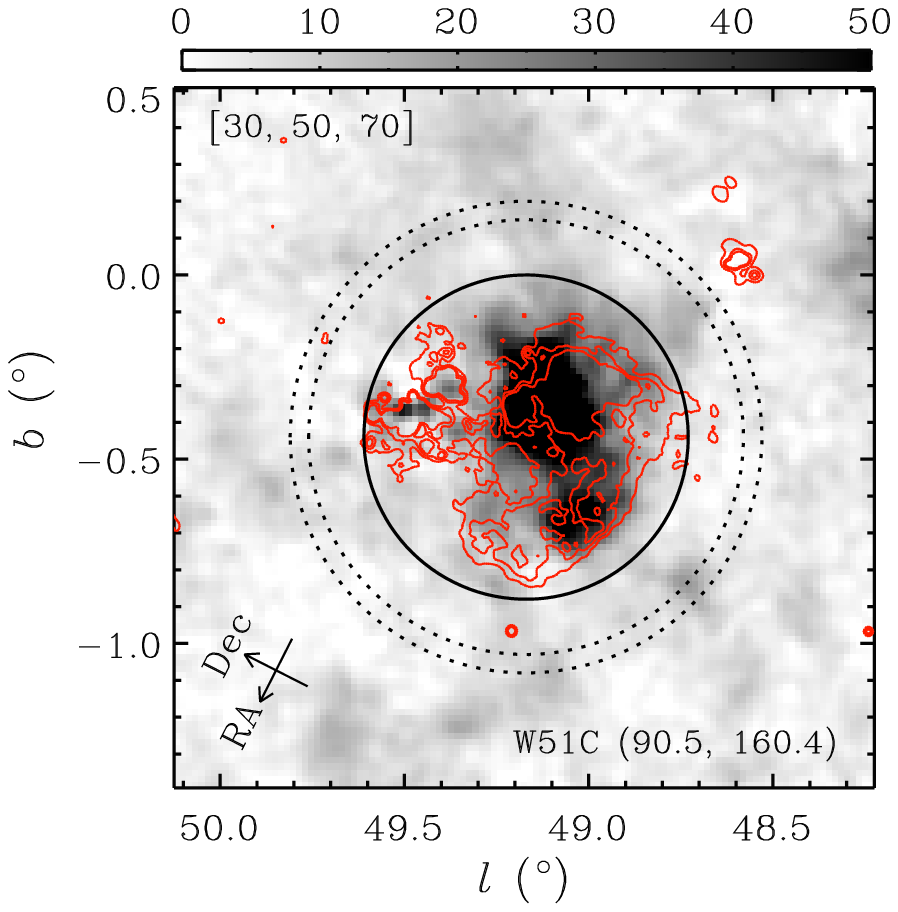}
\plottwo{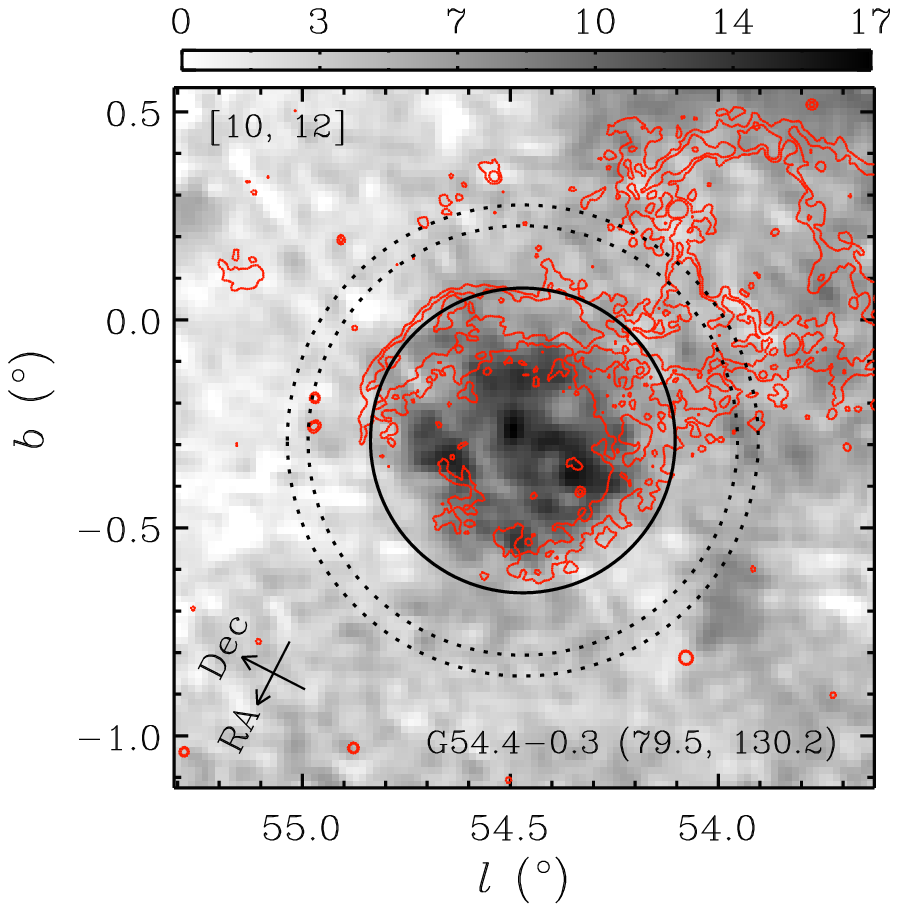}{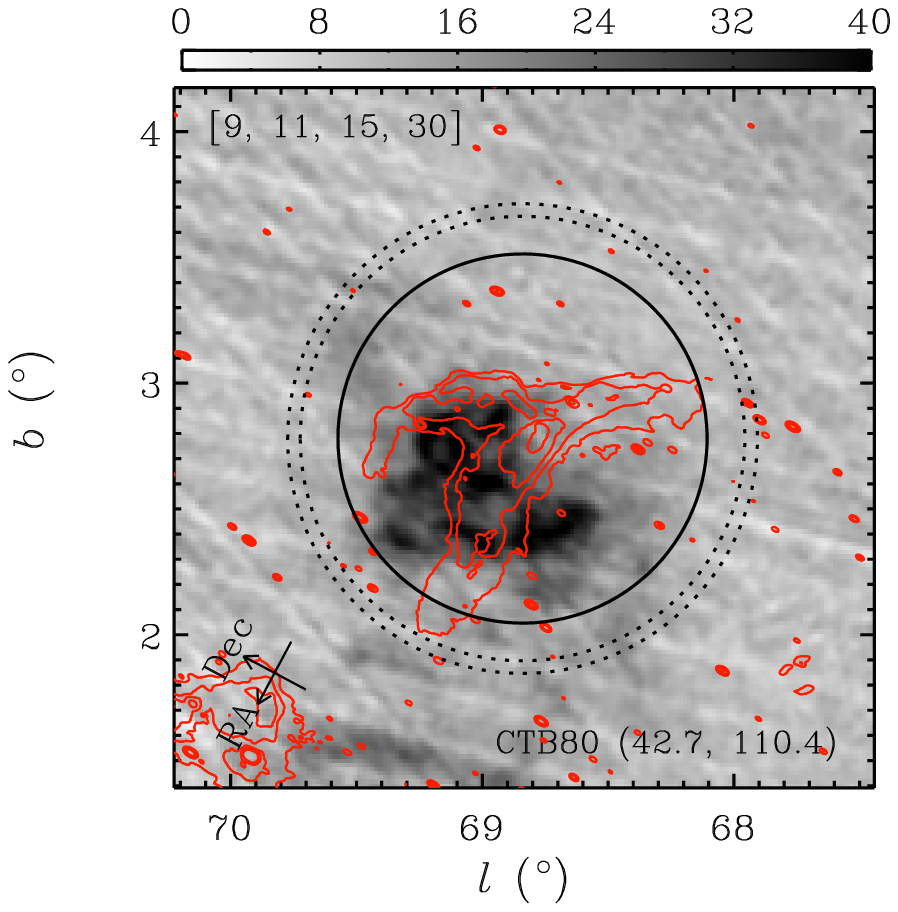}
\caption{
\schi maps of the fast-expanding \schi shells in four SNRs.
The lower right corner of each panel gives the SNR name and 
velocity range for integration in \kms.
For W44, both positive- and negative-velocity \schi emissions have
been used, although the former dominates.
The scale bar units are K~\kms.
Red contours show the SNR morphology in 21~cm radio continuum,
from the \cgps\ for CTB 80 and the \vgps\ for the others.
Contour levels are written at the upper left corner of each panel
in Kelvins.
The black solid circle in each plot
shows the mean size of the SNR in radio continuum.
Two dotted circles mark the annulus used for estimating the
background emission in deriving the average profiles in Figure~\ref{fig:vprof}.
The size of W51C is 48\arcmin,
and other sizes are from Green's catalog.
To aid comparison to literature studies in Equatorial coordinates, 
arrows are shown in each frame indicating right ascension and declination
(J2000) coordinate directions.
}
\label{fig:itg}
\end{figure}

We have identified four SNRs that have associated HV \schi gas: 
G$34.7-0.4$ (W44), G$49.2-0.7$ (W51C), 
\g54 (HC40), and G$69.0+2.7$ (CTB~80). 
All four were ranked~3 by KH91, and 
follow-up high resolution observations have been made for
SNRs W44, W51C, and CTB~80 \citep{koo95a, koo97a, koo93}.
Figure~\ref{fig:vprof} shows their average \schi line profiles 
while Figure~\ref{fig:lvmap} shows 
($\ell,v$) maps across the centers of these four SNRs. 
(See Figure~\ref{fig:itg} for the areas used to derive these profiles.) 
The average profiles show some fluctuations in baseline, but they are removed 
in background-subtracted profiles. 
Figures~\ref{fig:vprof} and \ref{fig:lvmap} 
show that these four SNRs have excess \schi emission at 
highest positive velocities. In W44, a faint, but clear, excess
emission at highest negative velocities is also visible.
Integrated ($\ell$, \itb) maps in Figure~\ref{fig:itg} 
show the spatial distribution of HV \schi emissions.
The overlaid radio contours of SNRs clearly show that 
the HV \schi gas is confined inside the SNR boundaries of 
W44 and \g54. 
We discuss W44 and \g54 
in some detail below, where new results are obtained.  We also briefly comment
on the other two SNRs, W51C and CTB~80, where the \igalfa\ results 
agree with previous findings.

\subsection{W44 (G$\bf34.7-0.4$)}
\subsubsection{Previous studies}

W44 is a middle-aged SNR ($\sim2$\ee4~yr) of mixed morphology, being
shell-type in radio continuum but centre-filled in X-rays \citep{rho98}.
The radio continuum shell is somewhat elongated ($35\arcmin \times 27\arcmin$), 
and the southeastern\footnote
{Directions in this paper are all in reference to 
Galactic coordinates, not J2000 Equatorial coordinates.}
portion of the SNR shows enhanced radio emission 
\citep[Figure~\ref{fig:itg}; see also][]{castelletti07}.
The pulsar PSR~1853+01 with 
a spin-down age of $\sim2$\ee4~yr lies
9\arcmin\ southwest of the center of W44,
embedded in an X-ray emitting
pulsar wind nebula \citep{petre02, wolszczan91}.
\schi gas at very high positive velocities 
($\simgt +130$~\kms) accelerated by the SNR shock has been detected 
inside the remnant and studied using the Arecibo telescope 
\citep{koo95a}. 
Ample evidence indicates the SNR is interacting with a 
molecular cloud in the southeast at $\vlsr=+46.6$~\kms\/
\citep[][and references therein]{seta98, reach05}. 
An extensive, organized system of thin and knotty \Ht\ filaments
filling the SNR 
may indicate that 
the SN exploded inside a molecular cloud \citep{reach05, froebrick11}.
Gamma-ray emission from the SNR has been detected by \fermi/LAT and 
\agile/GRID at 50~MeV---10~GeV
\citep{abdo10,giuliani11}. The emission is confined to an incomplete 
ring structure that matches well with the SNR but with a 
slight offset. The gamma-ray spectrum is well modeled with emission from 
cosmic-ray protons interacting with the nuclei in the ambient medium. 
Distance estimates to the SNR range from 2.6 to 3.2~kpc.

\subsubsection{New Results}
\label{sec:W44-newresults}

The \igalfa\ images reveal several new \schi features not 
seen in previous studies.
First, we detect fast-moving \schi gas associated with the SNR 
at both the highest positive and highest negative velocities.
The negative-velocity emission is relatively faint, 
but clearly real (see below).
The \schi gas at 
positive and negative velocities must be from the receding and approaching 
hemispheres of the SNR, respectively. 
{\em This is the first detection of both sides of an 
expanding SNR shell in the \schi 21-cm emission line.} 
We will discuss the visibility 
of fast-expanding \schi shells in \S~\ref{sec:stat}.

\begin{figure}[ht]
\epsscale{1.0}
\centering
\plottwo{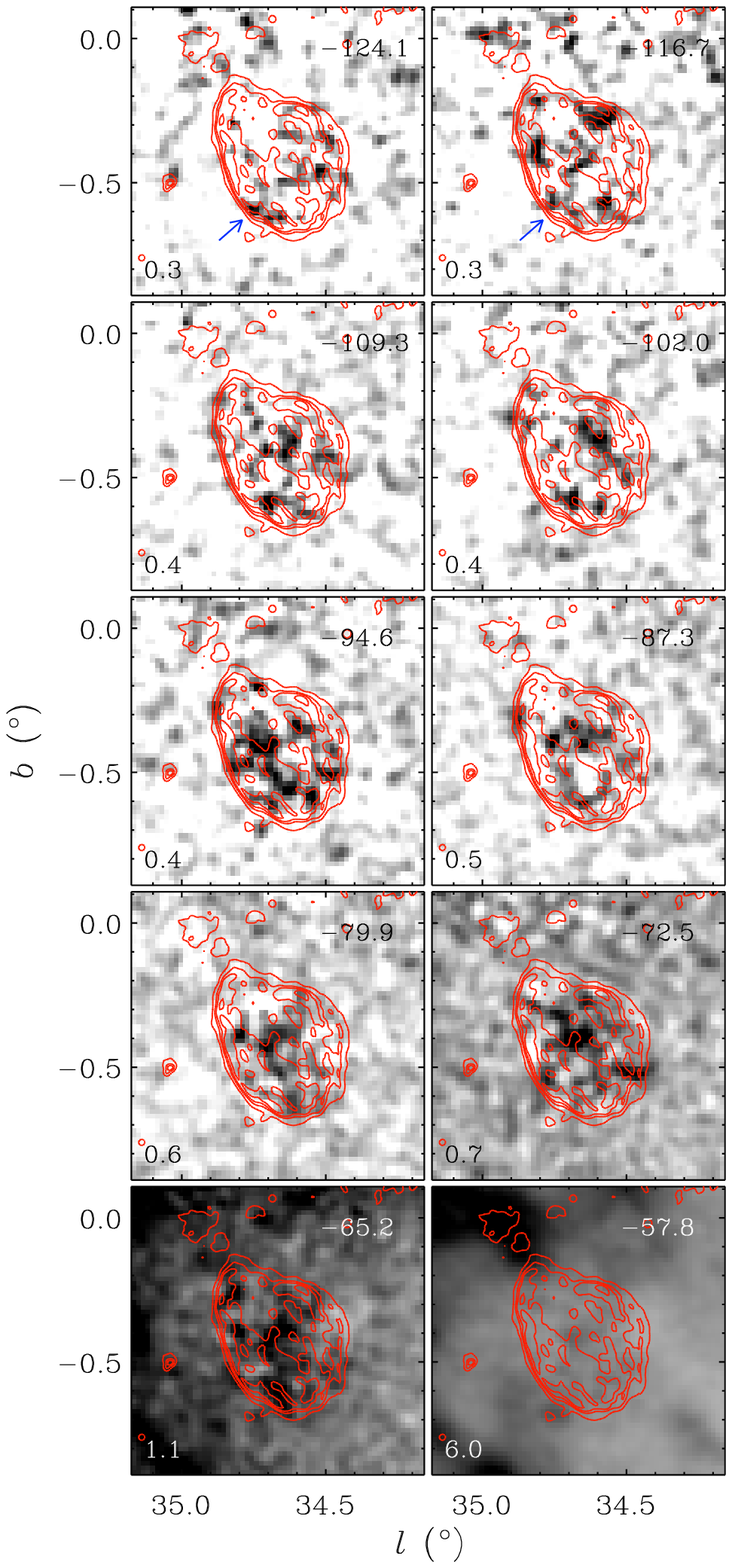}{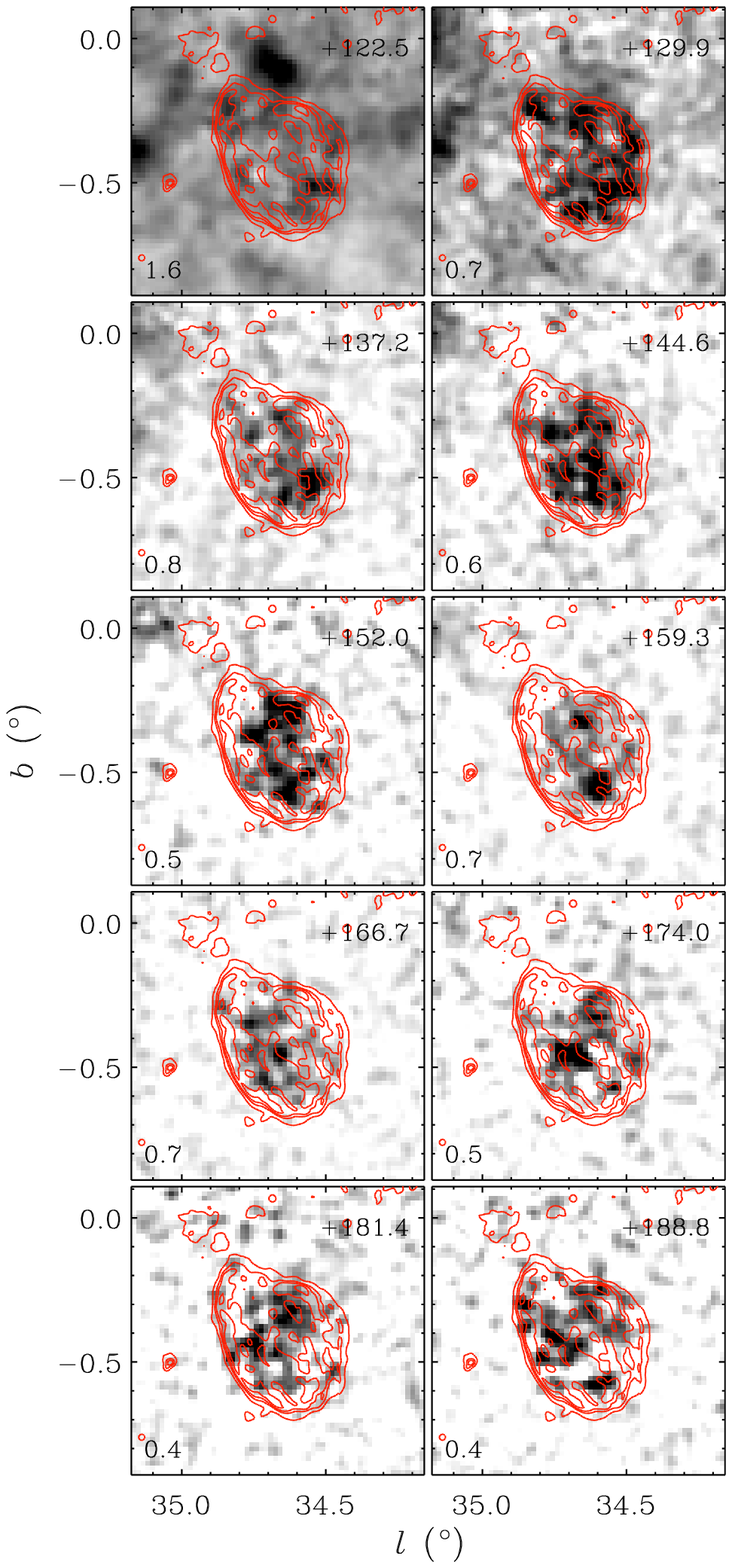}
\caption{
Channel maps of HV \schi gas associated with the SNR W44
at negative (left) and positive (right) velocities
in the \igalfa\ \schi data.
The central LSR velocity of each panel
is written at the upper right corner
in \kms.
The velocity width of one displayed channel is 3.68~\kms; 
this is binned by a factor of 20 from the raw data and has a measured RMS
noise of 0.07~K.
The \schi brightness temperature in each gray scale ranges from 
0~K (white) to the value (black) at the lower left corner of each panel 
in Kelvins. 
Radio continuum emission at 21~cm is shown
by red contours as in Figure~\ref{fig:itg}.
The blue arrows in the first two panels indicate
an emission feature described in \S~\ref{sec:W44-newresults}.
}
\label{fig:chmap-W44}
\end{figure}

Figure~\ref{fig:chmap-W44} shows channel maps 
at both negative and positive velocities 
where we see \schi emission confined inside the SNR boundary. 
At the highest negative velocities ($-120$ to $-100$~\kms), 
a knotty, elongated emission 
feature is seen along the southeastern boundary of the SNR
(blue arrow in first two panels). 
It is worth noting that this is where the SNR is currently interacting with the 
molecular cloud (see references in the previous paragraph). 
At velocities less negative than this, the emission in the central 
area has a ring-like shape.
At positive velocities, the \schi emission is more prominent.
At the highest positive velocities, the \schi gas is confined to the central 
area of the remnant with the ring-like morphology noted by 
\citet{koo95a}. And at $\vlsr=+130$ \kms, the 
\schi gas forms a ring structure along the inner boundary of the SNR. 
This indicates that the \schi gas is part of an expanding shell, and that 
the size of the \schi shell is comparable to 
the radio continuum shell.
This is in contrast to \citet{koo95a}, who concluded that the \schi shell 
was smaller than the continuum shell from 
the extrapolation of the \schi gas distribution at the highest velocities. 
Figure~\ref{fig:chmap-W44} indeed shows that the \schi ring structure 
rather abruptly shifts from the central region to the SNR boundary
as the velocity changes from $+137$ to $+130$~\kms. 
We speculate that the total extent of the expanding \schi shell is 
probably comparable to the radio continuum shell, but the shell is not complete.

\begin{figure}[ht]
\epsscale{0.7}
\plotone{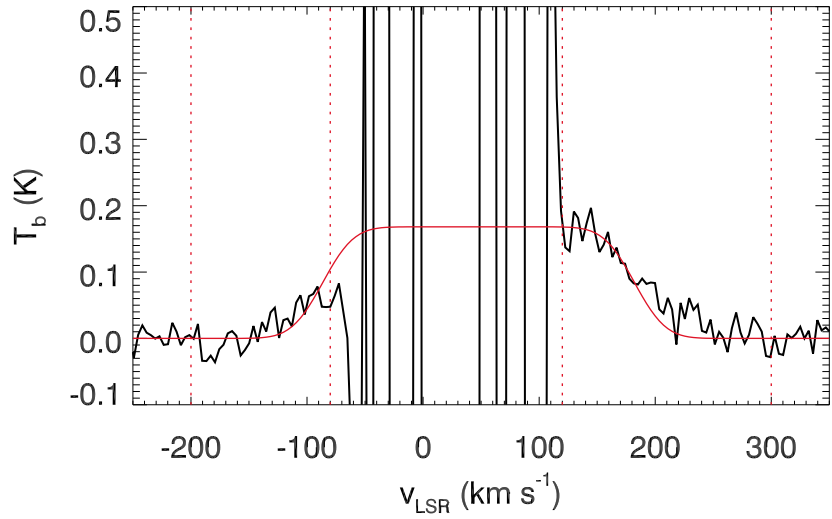}
\caption{
A fit to the average background-subtracted \schi profile of W44. 
Low-velocity residual brightness fluctuations are as in Figure~\ref{fig:vprof}.
The red solid line is a best fit to the profile.
The red dotted lines mark the velocity range where the fit has been performed.
See \S\ref{sec:W44-newresults} for an explanation of the fit.
}
\label{fig:vprof-W44}
\end{figure}

Figure~\ref{fig:vprof-W44} 
shows the average \schi profile of W44. 
If the \schi shell is symmetric, we can 
in principle derive an accurate systemic velocity by measuring the 
velocity at which the line profile becomes symmetric. 
For example, if we naively fit 
a Gaussian to the line profile, we obtain a central velocity of 
$+59$~\kms, which is considerably higher than 
the systemic velocity ($+47$~\kms) of the associated molecular cloud.
We attribute the discrepancy to the asymmetric 
mass distribution of the \schi shell. The implied asymmetry also makes it 
difficult to derive an accurate mass. But the uncertainty in the mass 
estimation is mainly due to the unobservable mass at low velocities, and,  
by fixing the central velocity, we can derive a reasonably accurate mass. 
The average line profile becomes flattened at velocities below
+160~\kms, i.e., it does not rise steeply as one might expect when the
high-velocity part ($\ge 160$~\kms) is the tail of a Gaussian profile,
which is consistent with \citet{koo95a}. 
If the \schi shell is uniform and
expanding at a constant speed $\vexp$, then the 
average \schi 21-cm line profile would appear 
as a square profile that extends 
from $-\vexp$ to $+\vexp$ centered at the systemic velocity $v_0$
\citep[e.g., see][]{koo91}, i.e.,
\begin{equation}
T_{b,\rm square}(v)=\begin{cases}
T_{b,max} & {\rm for}~v_0 -v_{\rm exp} \le v \le v_0 + v_{\rm exp}  \\
0, & {\rm otherwise}
\end{cases}
\end{equation}
where $T_{b,max}$ is the maximum brightness temperature, which is proportional 
to the total mass in the shell (see below). 
But the shell probably does not coast at a constant 
speed, given
non-uniformity/inhomogeneity of the ambient medium,  
turbulent motions produced by hydrodynamic instabilities, etc. 
We assume that the observed 
profile is composed of a flat profile convolved by a Gaussian profile, i.e.,
\begin{equation}
T_b(v)=\int_{-\infty}^{+\infty} 
T_{b,\rm square}(u)\exp\left[{-(v-u)^2} \over 2{\sigma_v}^2\right] du
\end{equation}
where $\sigma_v$ is the dispersion of random velocities related to 
the velocity at Full-Width at Half Maximum (FWHM) $\Delta v_{\rm FWHM}$ by 
$\sigma_v = \Delta v_{\rm FWHM}/2\sqrt{2\ln{2}}$. 

\begin{figure}[ht]
\epsscale{1}
\plotone{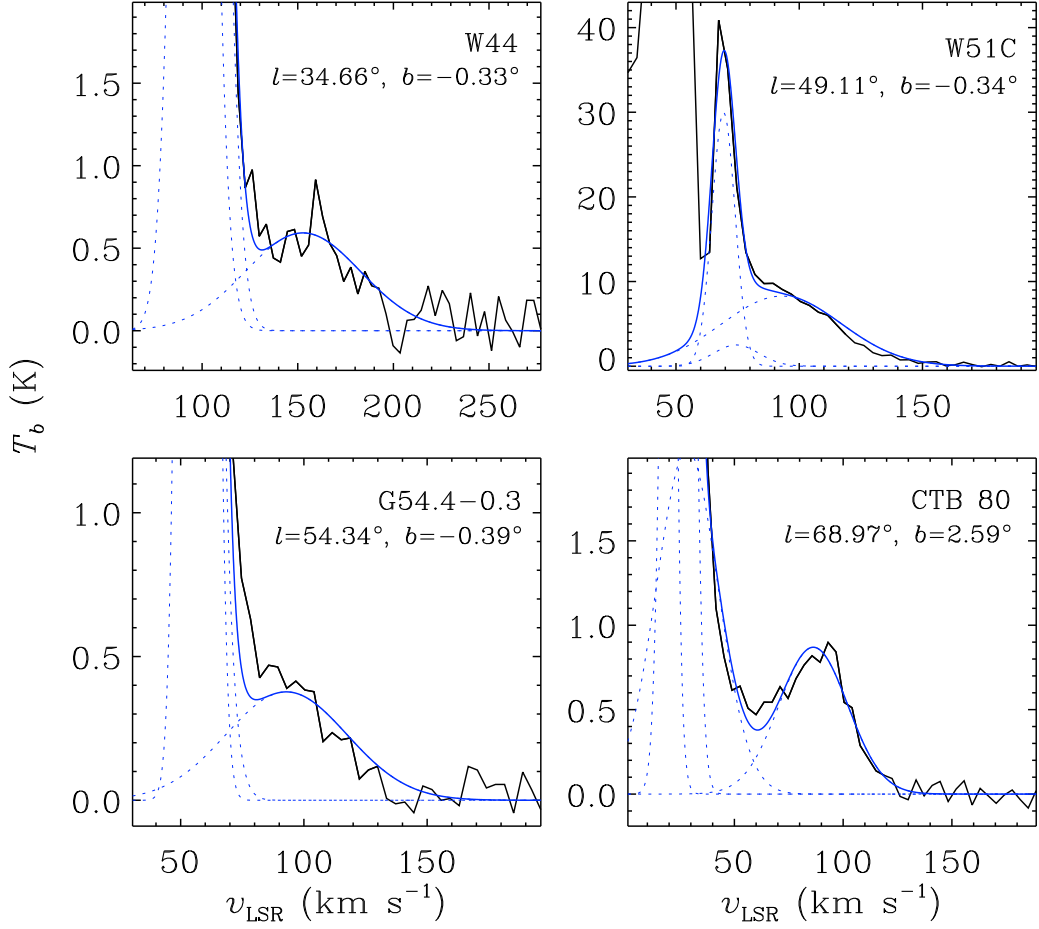}
\caption{
\schi 21-cm spectra of some prominent HV clumps in individual SNRs (black lines).
Their positions are given in the upper right corner of each panel.
The HV parts of the spectra have been fitted by several Gaussian components, and
the blue dotted and solid lines show the profiles of individual components and their
sum, respectively. The components at the highest velocities are the ones
related to the fast-expanding \schi shells.
}
\label{fig:vprof-repr}
\end{figure}

The fast-moving gas is clumpy and has broad lines.
In the SNR W51C, for example, 
\citet{koo97a} resolved individual clumps using the 
VLA and found that their line profiles have FWHM of 
$\sim 40$~\kms. In the \igalfa\ survey, we could not 
completely resolve
out clumps from the background emission 
in velocity space. But we see many clumps that appear as 
prominent protrusions. Figure~\ref{fig:vprof-repr} shows some 
examples of HV bumps in the tails of 
strong background emission.
We fit the HV parts of these spectra  
with several Gaussian components. 
Our purpose is 
only to obtain the widths of the HV bumps, 
so the fit parameters of the low-velocity components are 
unimportant. 
We obtain suitable results using 
four components for CTB~80 and three for the others (Figure~\ref{fig:vprof-repr}),
and we find $\Delta v_{\rm FWHM}\approx 50$~\kms\ 
for the HV components in all four SNRs.
We adopt $\Delta v_{\rm FWHM}=50$~\kms\ as the characteristic value for 
the fit. 
The large velocity dispersion within the shell implies that 
the shell is dispersing, but 
its thickness will remain small compared to the radius of
the shell until the very late stage of SNR evolution.

With the systemic velocity fixed at $v_0 \equiv +47$~\kms, we vary 
$T_{b,max}$ and $\vexp$. The fit was done 
using the IDL routine MPFIT \citep{markwardt09}.
The best fit profile is shown in Figure~\ref{fig:vprof-W44}. 
Its parameters are 
$\vexp=135\pm 2$~\kms\ and $T_{b,max}=0.168\pm 0.005$~K.
The \schi mass is derived from the average column density 
obtained from the fit,  
i.e., $\bar N_{\rm H}=1.822\ee{18}~T_{b,max}\times 2\vexp$~cm$^{-2}$
considering both approaching and receding sides of the shell, 
by multiplying the area of W44. 
(Since the peak brightness temperature is
only $\sim 0.2$~K, we may assume that the \schi emission is optically thin.)
The derived \schi mass is $393\pm 13$~\msol, 
which implies a kinetic energy of $9.9 \pm 0.4$\ee{49}~erg.
These numbers agree with those of \citet{koo95a} within 10--20\%. 
The resulting 
dynamical age of the shell is $t\simeq 0.3 R_s/v_s\simeq 2.7\ee4$~yr
\citep{cioffi88},
where we used a geometrical mean radius (12.5~pc at 2.8~kpc;
See Tables~\ref{tab:rank3} and \ref{tab:distance} 
for the adopted distances to the SNRs) 
as the radius of the SNR, $R_s$.
This dynamical age is somewhat larger than 
the characteristic spin-down age of the pulsar.  
Assuming that this mass was initially inside the volume of the SNR, 
i.e., inside a sphere of geometrical radius of $12.5$~pc, the mean 
density of hydrogen nuclei 
in the ambient medium would be $1.9\pm0.1$~cm$^{-3}$. 
The initial explosion energy, $E_{\rm SN}$, can be estimated from
\begin{equation}
E_{\rm SN}=6.8\times 10^{43} \, {n_0}^{1.16} \, {R_s}^{3.16} \, {\vexp}^{1.35}~~~{\rm erg}, 
\end{equation}
where $n_0$ is the ambient density of hydrogen nuclei
in cm$^{-3}$, $R_s$ is in pc, and $\vexp$ is in \kms\ \citep{cioffi88}. 
We assume solar metallicity.
Substituting the derived values, i.e., 
$n_0=1.9$~cm$^{-3}$, $R_s=12.5$~pc, and $\vexp=+135$~\kms, 
we have $E_{SN}=3.2\pm 0.1 \ee{50}$~erg.
This is the energy required to heat and accelerate 
more or less uniformly distributed atomic gas. 
Since there is also fast-expanding ($\sim+30$~\kms), dense 
molecular gas accelerated by SN blast wave \citep{reach05, froebrick11}, 
some fraction of the total SN energy should have been 
used up for heating and accelerating the molecular gas.
It would be interesting to derive the kinetic energy of 
shocked \Ht\ filaments and compare it to that of the \schi shell. 

\subsection{G$\bf54.4-0.3$ (HC 40)}

\subsubsection{Previous studies}

\g54 is a shell-type SNR of circular shape with a diameter of 
40\arcmin. The shell is not complete, and its radio continuum 
brightness is not uniform
\citep[see the contour map in Figure~\ref{fig:itg};][]{junkes92a}.
In the southeastern part of the remnant, the circular portion of 
the shell is missing, and instead faint continuum structures are visible 
connected to the tips of the bright ends. 
This probably indicates that the SNR blew out 
in this direction, presumably on encountering a rarefied medium there, 
and the shell was disrupted in the process. 
\citet{junkes92a} carried out CO observations and suggested an association with 
a molecular cloud at $\vlsr = +36$ to $+44$~\kms. 
They proposed a distance
of 3~kpc using the rotation curve of \citet{burton88}. 
\cite{case98} adjusted this distance by adopting a revised rotation
curve with $R_\odot = 8.5$~kpc and $\Theta_\odot = 220$~\kms\ 
and obtained 3.3~kpc.  
If we use the rotation curves of \cite{brand93} or \cite{levine08}, 
the near-side kinematic distance corresponding to the systemic velocity of +40~\kms\ 
in this direction is 3.9 and 3.1~kpc, respectively.
We adopt the 3.3~kpc distance of
\citet{case98} in this paper.
There are several \schii regions in this area at about the same 
distance, including the compact \schii region just outside
the northwestern boundary of the remnant \citep{junkes92b}.
\citet{junkes96} observed the remnant with \rosat\ and 
derived an absorbing hydrogen column density of 
1\ee{22}~cm$^{-2}$ and a plasma temperature of 2\ee7~K.
\cite{boumis05} detected optical emission lines in the northwestern edge of 
the shell and derived an absorbing hydrogen-nuclei 
column density of 2.9--4.0\ee{22}~cm$^{-2}$.
KH91 detected fast-moving \schi gas at velocities $+91$ --- $+108$~\kms.

\subsubsection{New Results}
\label{sec:g54-newresults}

\begin{figure}[ht]
\epsscale{1.0}
\plotone{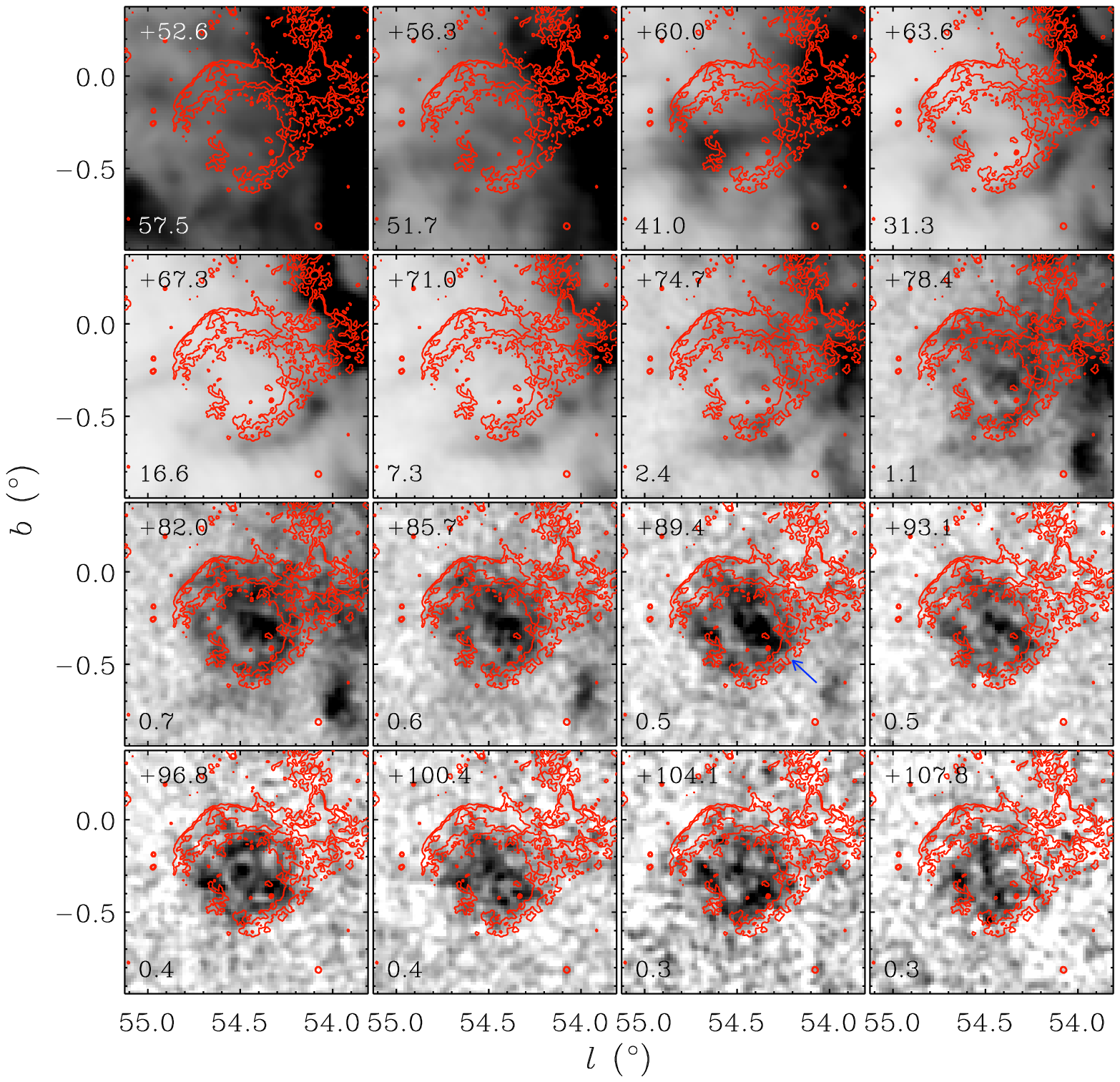}
\caption{
Channel maps of HV \schi gas associated with the SNR \g54
in the \igalfa\ \schi data. 
As in Figure~\ref{fig:chmap-W44}, the central LSR velocity of each panel
is shown in the upper left corner
in \kms,
and the velocity width of one channel is 3.68~\kms.
The \schi brightness temperature in each gray scale ranges from
0~K (white) to the value (black) at the lower left corner of each panel
 in Kelvins.
Radio continuum morphology of the SNR is shown by red contours
as in Figure~\ref{fig:itg}.
The blue arrow in the +89.4~\kms\ panel marks
an emission feature described in \S~\ref{sec:g54-newresults}.
}
\label{fig:chmap-HC40}
\end{figure}

This SNR has some of the
most prominent \schi shell structure in our sample.
Figure~\ref{fig:chmap-HC40} shows \schi channel images of \g54 at
$\vlsr= +53$ to $+108$~\kms.
At the highest positive velocities, e.g., $\vlsr= +100$~\kms, 
the \schi gas is 
confined to the central area of the remnant, whereas at 
lower velocities, we see a well-defined ring structure
as well as some emission filling the central area. 
This velocity structure indicates that the HV gas is the 
receding portion of an expanding shell. 
The shell has non-uniform brightness.
A prominent feature is the bar-like structure extending 
from the center to the southwestern at $\vlsr= +78$ to $+93$~\kms\
(blue arrow in a panel at $+89$~\kms).
Its northeastern tip appears as a bright spot 
at $(54\dotdeg35,-0\dotdeg30)$ at $\vlsr= +82$ --- $+89$~\kms.
At low velocities, e.g., at $\vlsr= +64$ --- $+78$~\kms, 
we see some correspondence between \schi and 
radio continuum structures. First, the same southeastern portion of the 
\schi shell is also weak and missing as in the continuum shell. Second, 
there is faint but enhanced \schi emission
coincident with the continuum shell at $\ge 70$~\kms.
Third, along the southwestern SNR shell, there is a larger filamentary 
\schi structure just outside it at 
$\vlsr= +64$ to $+78$~\kms.  This external \schi feature has the
same curvature as the SNR shell, so it might be 
associated with the SNR too.  
Perhaps this is the boundary of the 
stellar wind bubble produced by the progenitor star,
although it is not obvious how we can see the SNR shell 
inside a wind bubble.

\begin{figure}[ht]
\epsscale{0.7}
\plotone{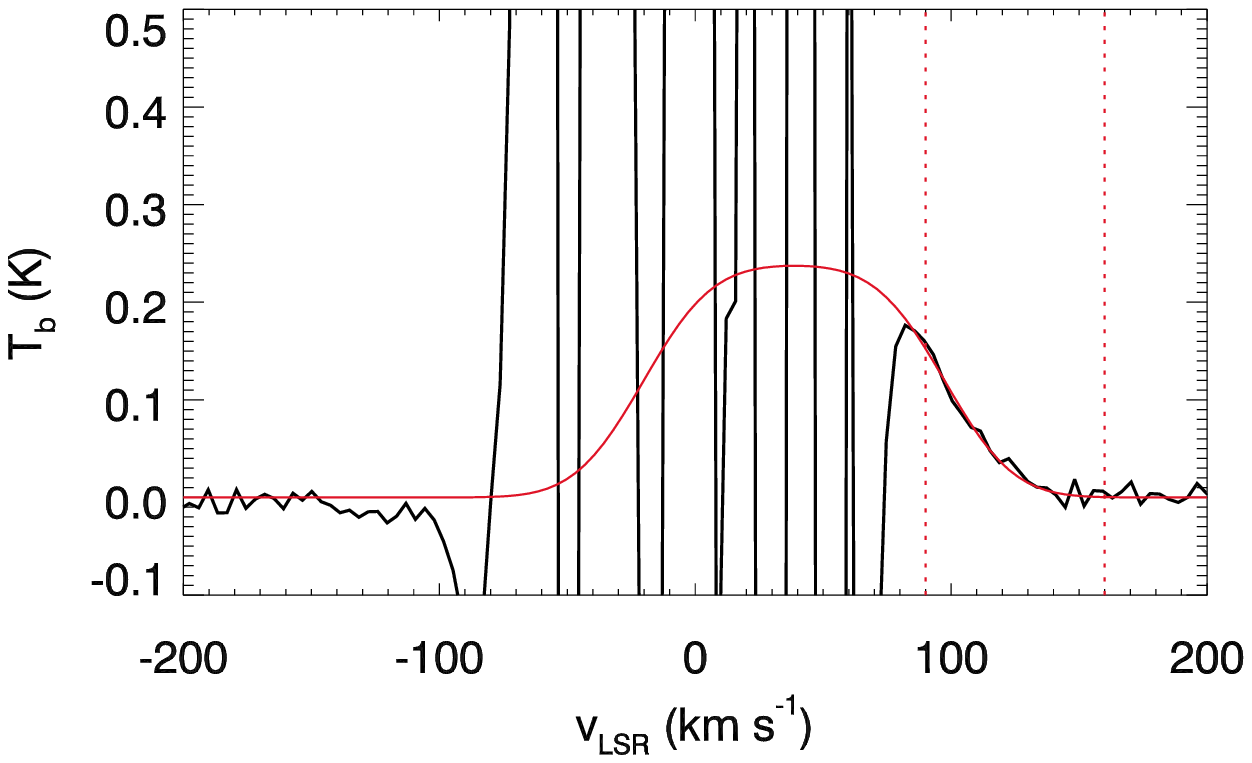}
\caption{
A fit to the average \schi profile of \g54.
Low-velocity residual brightness fluctuations are as in Figure~\ref{fig:vprof}.
The red solid line is a best fit to the profile.
The red dotted lines mark the velocity range where the fit has been performed.
See \S\ref{sec:W44-newresults} for an explanation of the fit.
}
\label{fig:vprof-HC40}
\end{figure}

We derive the mass of the G$54.4-0.3$ \schi shell 
in a similar fashion to W44, but using only positive 
velocities. We adopt $\Delta v_{FWHM}=50$~\kms\ and 
assume a systemic velocity of $+40$~\kms.
The best-fit profile, shown in Figure~\ref{fig:vprof-HC40}, uses
$\vexp=59\pm 6$~\kms\ and an \schi mass of $580\pm 150$~\msol\
at $3.3$~kpc. 
The corresponding kinetic energy is $2.8 \pm 0.9 $\ee{49}~erg.
The resulting 
dynamical age of the shell is $t\simeq 0.3 R_s/v_s\simeq 9.5\ee4$~yr,
where we used $R_s=19.2$~pc (20\arcmin).
Again assuming that this mass was initially inside the volume of the SNR, 
the mean density of hydrogen nuclei 
in the ambient medium would be $0.79\pm 0.20$~cm$^{-3}$. 
With the above parameters, we obtain $E_{SN}=1.5\pm 0.5 \ee{50}$~erg. 
This is considerably smaller than 
the canonical value of 1\ee{51}~erg but not unreasonable.

\subsubsection{Others}

\noindent
G$49-0.7$ (W51C) is a middle-aged, shell-type SNR interacting 
with a molecular cloud.
\citet{koo97a, koo97b} obtained
high-resolution 
\schi and CO observations and developed a model in which the 
fast-moving \schi gas is produced by the SNR shock propagating into a molecular
cloud. The SNR is one of the most luminous $\gamma$-ray sources in the Galaxy
\citep{abdo09}. The \igalfa\ results for the HV \schi gas 
in Figures~\ref{fig:vprof}--\ref{fig:itg} agree with previous results.

\noindent
G$69.0+2.7$ (CTB~80) is one of the first infrared SNRs detected by \iras\
\citep{fesen88}. It appears as a large ($\sim1$\arcdeg), spherical shell-type 
SNR in IR while, in radio continuum, 
only the northern portion of the shell is bright 
due to the interaction with the pulsar. \citet{koo90, koo93} carried out \schi 
studies using the Arecibo telescope and the VLA and confirmed the large size 
of the SNR shell and its old age ($\sim1$\ee5~yr). 
Again, the \igalfa\ HV \schi results in Figures~\ref{fig:vprof}--\ref{fig:itg}
agree with previous results.

\section{Discussion}
\subsection{Properties of \schi SNRs}

We have detected 4 SNRs with fast expanding \schi shells
in the \igalfa\ survey area. 
Table~\ref{tab:rank3} summarizes their parameters: 
distance $d$, radius $R_s$, systemic velocity $v_0$, 
expansion speed $\vexp$, dynamical age, 
\schi mass, kinetic energy, ambient density $n_0$, and 
initial explosion energy $E_{\rm SN}$. The table also lists whether the remnant
has an associated pulsar and whether it is interacting with a molecular cloud.
The parameters of W44 and \g54 are those derived in this work,
whereas those of W51C and CTB~80 are from previous studies. 

There are several points to make.
First, all 4 SNRs are middle-aged (1.8--9.5\ee4~yr).
Second, the ambient densities are $\simgt 1$~cm$^{-3}$, 
considerably larger than the densities of either the warm 
or hot diffuse ISM filling most of the interstellar volume.
In particular, three SNRs are interacting with molecular clouds.
Third, two SNRs are the remnants of core-collapse SNe with associated 
pulsar wind nebulae (PWNe). 
The other two
remnants, \g54 and W51C, do not have associated PWNe, 
but their interactions with molecular clouds suggests
they also likely have massive progenitors.
In summary, the SNRs with \schi shells (hereafter \schi SNRs) that 
are detected are middle-aged SNRs of probable CCSN origin 
interacting with a relatively dense medium.

The dynamical evolution of middle-aged SNRs in a uniform medium 
was studied in detail by \citet{cioffi88}, who derived    
analytic expressions for radius $R_s$ and expansion velocity $v_s$
that describe the results of their one-dimensional numerical simulations. 
\citet{koo04a} proposed somewhat simpler but still accurate forms of 
the equations of \citet{cioffi88}.\footnote{ 
There was a typo in equation (8) of \citet{koo04a}: the index 
``-1/14'' should be read as ``1/14''.
}
We find it useful to introduce a parameter 
$\delta\equiv n_0 {E_{51}}^{-0.861}$ where 
$n_0$ is the density of hydrogen nuclei in the ambient medium 
divided by 1 cm$^{-3}$, and 
$E_{51}$ is the SN energy released to the ISM in units of 
$10^{51}$~erg. Note that $\delta$ is dimensionless.
The advantage of introducing $\delta$ is that the radius and velocity 
of middle-aged SNRs are now related by 
\begin{equation}
R_s=25.7 \delta^{-0.367} {v_{s,2}}^{-3/7}~~~{\rm pc},
\end{equation} 
where $v_{s,2}\equiv (v_s/100~\kms)$. This equation is obtained
by combining equations (5)--(8) of \citet{koo04a}. 
(Note that this 
is just another expression of equation (3) except that their numerical 
coefficients differ by 2\%.)
In the $(R_s,v_s)$ plane, \schi SNRs evolve along a line satisfying the 
above equation. 
In terms of $\delta$, the expansion velocity of \schi 
SNRs is given by \citep[Eq. 6 of ][]{koo04a}
\begin{equation}
v_{s} = 179 \, \delta^{1/7} \, {E_{51}}^{0.194}
\left({11\over 10} {t\over \tsf} -{1 \over 10}\right)^{-7/10}~~~{\kms}, 
\end{equation}
where the ``shell formation time" $\tsf$ representing the onset of 
the formation of \schi shell is  defined by \citep{cioffi88}
\begin{equation}
\tsf\equiv 3.61\times 10^4 \, {n_0}^{-4/7} {E_{51}}^{3/14}~{\rm yr} \;\; .
\end{equation} 
As pointed out by \citet{koo04a}, the maximum disagreement between 
the equations (4)--(5) and the original
equations (3.22)--(3.23) of Cioffi et al.\ is less than 0.1\% 
during $t \sim (1 - 13) \, \tsf$.
The radius and velocity are now functions of $\delta$, $E_{51}$, 
and $t/\tsf$ instead of $n_0, E_{51},$ and $t$. 
And since the dependence of $v_s$ on $E_{51}$ is weak, 
a single grid can be drawn in the ($R_s, v_s)$ plane to describe the 
evolution of \schi SNRs in different physical environments.  

\begin{figure}[ht]
\epsscale{0.7}
\plotone{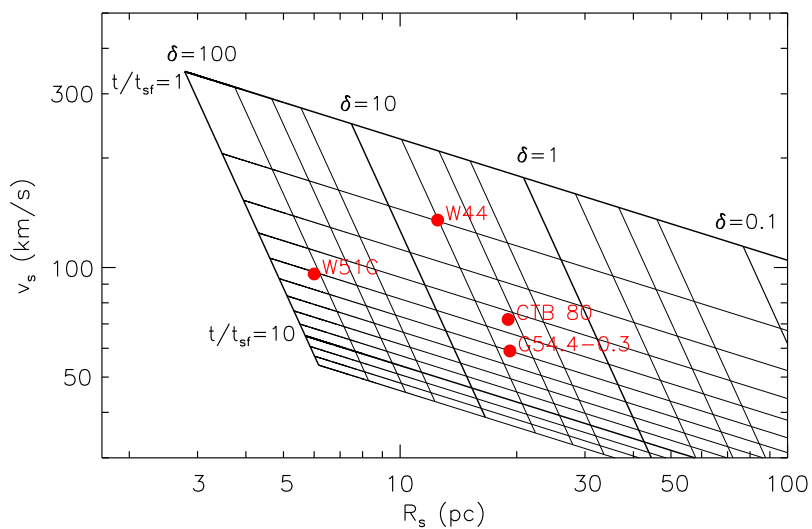}
\caption{
Radius-velocity relation of \schi SNRs. 
The grid is for $E_{51}=1$. The grid
shifts along lines of constant $\delta$
for different $E_{51}$ but not much: e.g.,
for $E_{51}=0.1$, the lines of constant $t/\tsf$ shift down by 36\%. 
The thin grid lines in $\delta$ are drawn at 20, 30, and 50\% of the thick-grid intervals, 
while those in $t/\tsf$ are at every 10\%. 
The four \schi SNRs identified in the \igalfa\ survey are marked.
}
\label{fig:snrs-rv}
\end{figure}

Figure~\ref{fig:snrs-rv} shows how the radius and velocity of \schi SNRs evolve 
in time ($t/\tsf$) for a given $\delta$.
The SNR develops a fast-expanding \schi shell 
at $t/\tsf=1$, which expands and slows down along a line of constant $\delta$
as it evolves.
For example, suppose an SNR has $E_{51}=1$ in a uniform medium with $n_0=1$, 
so that $\delta=1$. Then the SNR has an \schi shell of 
$(R_s,v_s)$ = (20~pc, $+179$~\kms) at $t/\tsf=1$ 
and $(R_s,v_s)$ = (40~pc, $+34$~\kms) at $t/\tsf=10$,
where $\tsf=3.61\ee4$~yr.  
The \schi SNRs identified in the \igalfa\ survey 
have radii of 6--19~pc and expansion velocities $+59$ --- $+135$~\kms, or 
$\delta=5$--50 and $t/\tsf=2$--9. 
The expansion velocities of the detected \schi shells are 
all greater than $+50$~\kms, which is necessary to be clearly discernible 
from the Galactic background emission (see next section).
It is also worth noting that 
no large HI SNRs expected in the diffuse ISM have been
detected, i.e., there are no \schi SNRs where $\delta<1$ in Figure~\ref{fig:snrs-rv}.
This may be either because such SNRs are rare or because such
SNRs are faint in radio continuum and
``missed'' in the current catalog of SNRs. We will discuss this further
in next section.

\subsection{Visibility and Statistics of \schi SNRs}
\label{sec:stat}

The visibility or detectability 
of SNRs in the \schi 21-cm line was investigated by \citet{koo04a}. 
An important constraint on the visibility of 
\schi SNRs is that they should be in the ``right'' positions
in the Galaxy where the line-of-sight velocities
of expanding \schi shells
can easily exceed the maximum or minimum LSR velocities 
of the Galactic background \schi emission,
e.g., along the loci of tangential points in the inner Galaxy. 
For example, W51C is 
in a high-visibility location,
because its systemic velocity
($+62$~\kms) is close to the maximum velocity ($\sim+90$~\kms; 
see Figure~\ref{fig:lvmap}) 
of the background emission in this direction ($\ell=49\dotdeg2$).
In contrast, W44 is 
in a low-visibility location: 
its systemic velocity
($+47$~\kms) is much less than the maximum 
velocity ($+130$~\kms) in this direction ($\ell=34\dotdeg7$).
Apparently, W44 is relatively young and 
has the largest expansion velocity, so that 
both its approaching and receding parts can be seen.

\begin{figure}[ht]
\epsscale{1.35}
\plottwo{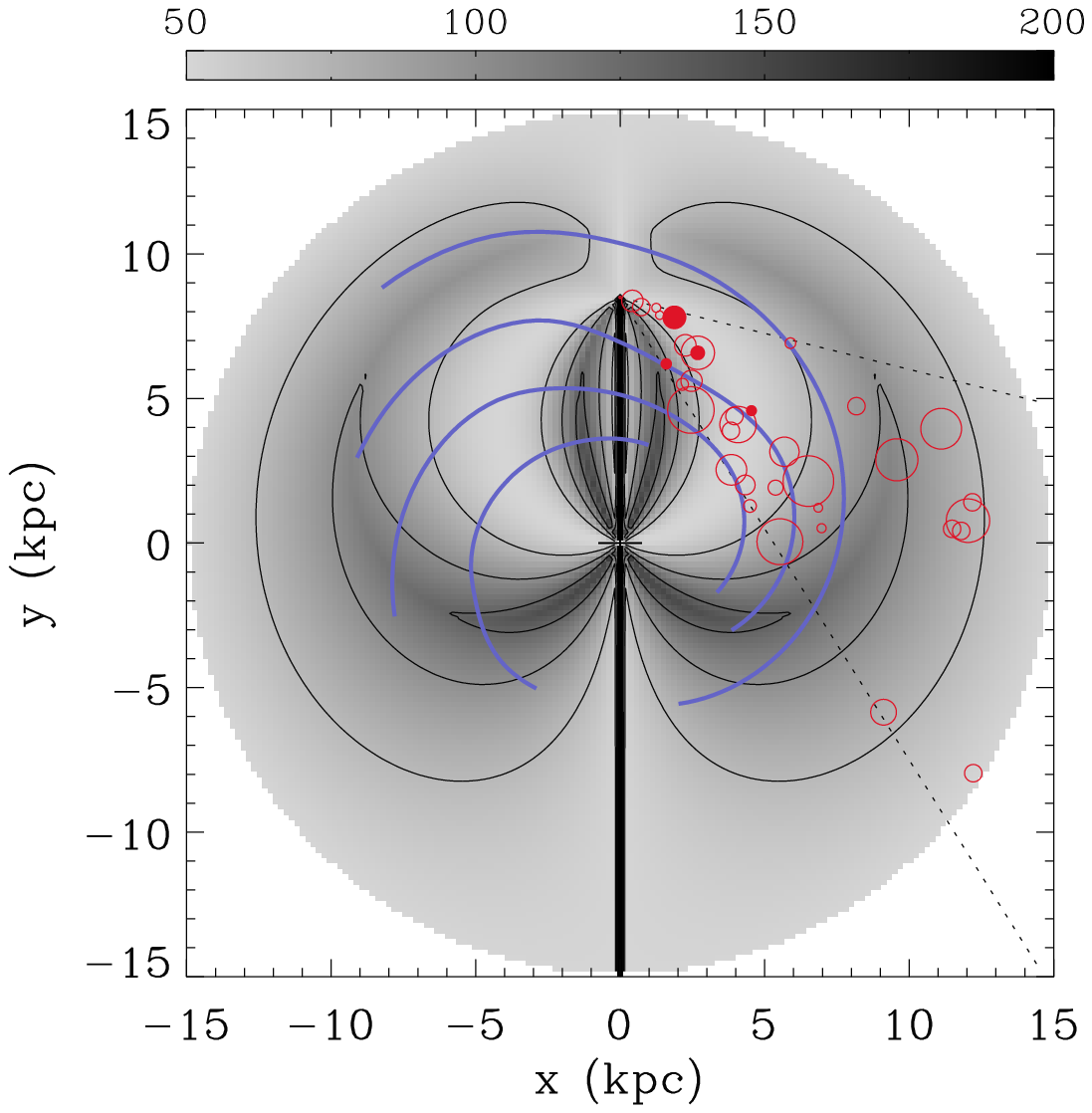}{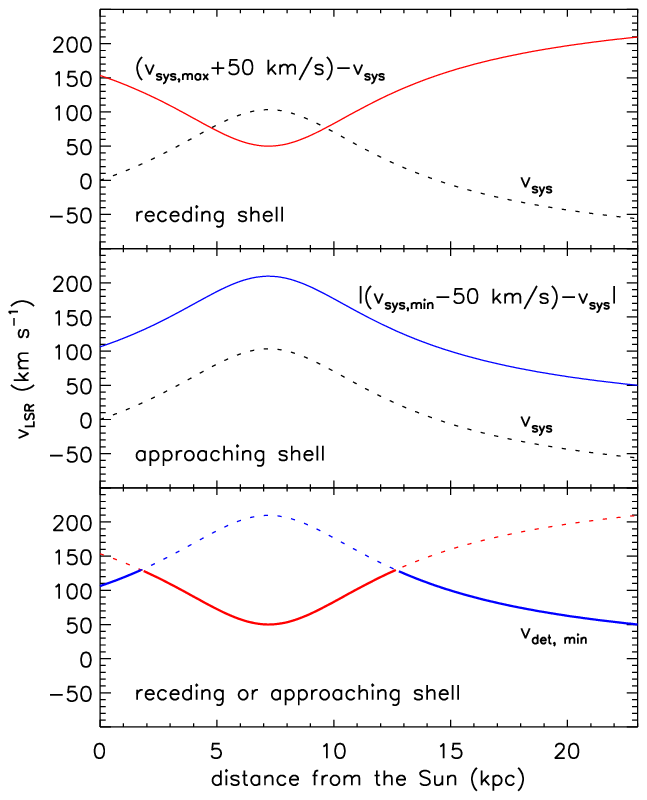}
\caption{
(left) Distribution of \vdet, the minimum expansion velocity of a 
fast-expanding SNR \schi shell for detection.
Areas with higher minimum expansion velocities  
are darker. The scale bar at the top displays velocity scales in a unit of \kms.
Contour levels are drawn at 70, 100, and 130 \kms.
The Sun is located 
at $x=0.0,y=8.5$~kpc. The dotted lines mark the boundaries
of the \igalfa\ survey at $b=0\arcdeg$, i.e., $\ell=32\arcdeg$ to $77\arcdeg$.
The blue curved lines represent the four spiral arms of \citet{taylor93}. 
The locations of the 
SNRs in the \igalfa\ area are marked by red circles with 
diameters proportional to SNR size.
The SNRs with fast-expanding \schi shells are marked by 
the filled circles. (right) One-dimensional velocity profiles at $\ell=32^\circ$.
Minimum expansion velocities required for the detection of receding (top frame) and
approaching (middle frame) portions of the shell are shown together with 
the systemic LSR velocity ($v_{\rm sys}$) as
a function of distance from the Sun. \vdet\ is the smaller
of the two velocities (bottom frame). (See text for details.)
}
\label{fig:snrs-gp-wb}
\end{figure}

The locations of the SNRs in the \igalfa\ area are marked 
in Figure~\ref{fig:snrs-gp-wb} (left), 
with filled circles for the 4 \schi SNRs.
Table~\ref{tab:distance} lists our adopted distances,
where the fifth column gives 
the distances that are considered to be reliable. 
Twenty-eight SNRs have reliable distances, half of which 
are from the compilation by \citet{green09a}. The 
other half are from our own literature search, with references
listed in the last column.
The sixth column gives distances estimated using the 
surface brightness -- diameter ($\Sigma-D$) relation
\citep[e.g.,][]{case98,arbutina04,guseinov03}.
For this work, we adopt the \citet{case98} version.
The $\Sigma-D$ relation has  
considerable dispersion, and its applicability 
has been criticized \citep[e.g.,][]{green04}. But 
without any other estimates, it still provides a useful reference.
We use reliable distances wherever possible and $\Sigma-D$ distances
for other SNRs.

The background grey-scale map in Figure~\ref{fig:snrs-gp-wb} (left) shows the 
minimum shell expansion velocity for detection, \vdet.
The Galactic disk is assumed to be  
axi-symmetric with radius 15~kpc and a flat rotation curve 
with $R_\odot = 8.5$~kpc and $\Theta_\odot = 220$~\kms. 
An expanding \schi shell is considered to be visible if 
the LSR velocity of its receding endcap, 
which is the sum of the systemic LSR velocity 
and the expansion velocity, 
is greater than the maximum LSR velocities
allowed by circular rotation 
in that direction by more than 50~\kms~or vice versa.
This appears to be conservative considering
that the turbulent velocity dispersion of the warm neutral medium 
is 27 \kms\ \citep{heiles03} and that the non-circular velocities due to
spiral shocks are typically 20~\kms\ \cite[e.g.,][]{roberts69}. 
But note that the
\schi emission from fast-expanding SNR shells appears as very weak
broad wings superposed on the Gaussian tails of background emission;
in Figure~\ref{fig:vprof-repr}, for example, the
maximum LSR velocities in the directions of the four SNRs according
to the flat rotation curve are,
in the order of increasing Galactic longitude,
 +94, +54, +42, and +15~\kms, respectively.
\schi shells along the tangent points in the inner Galaxy have small 
\vdet\ ($\sim50$~\kms) because their positive-velocity wings can 
be easily detected. 
The \schi shells near the outer boundary
of the disk in the survey area also have small \vdet, 
but, in this case, it is because
their negative velocity wings can be easily detected.
In order to help the understanding, 
Figure~\ref{fig:snrs-gp-wb} (right) shows the variation of 
\vdet\ for SNRs in the direction at $\ell=32^\circ$. 
In the top and middle frames, the dotted line shows how the systemic 
LSR velocity ($v_{\rm sys}$) 
varies with distance from the Sun due to Galactic rotation. 
Note that the maximum and minimum systemic LSR velocities in 
this direction are 103 and $-$56~\kms, respectively. 
This gives an approximate velocity range of the 
background \schi emission. 
Therefore, for the receding portion of an expanding SNR shell 
to be detectable, its expansion velocity 
should be larger than $(103+50)$~\kms$-v_{\rm sys}$,  
where $v_{\rm sys}$ is the systemic velocity of the shell 
(red line in the top frame).
On the other hand, for the approaching portion to be detectable, 
which will appear as a negative-velocity wing, 
the expansion velocity 
should be larger than $|(-103-50)$~\kms$-v_{\rm sys}|$       
(blue line in the middle frame). 
For just one part of the shell detectable, the required 
expansion velocity will be the smaller of the two 
(thick line in the bottom frame).
Figure~\ref{fig:snrs-gp-wb} shows that three SNRs (W51C, \g54, CTB~80) are located
where the \vdet\ is relatively small, i.e., 
50~\kms, whereas W44 is located
in a region where \vdet\/~$\sim90$~\kms. 
Table~\ref{tab:distance} lists \vdet\ at the position of each SNR.

In Figure~\ref{fig:snrs-gp-wb},  there are many 
SNRs that are {\em not} detected in \schi 21-cm line in spite 
of their favorable locations in areas where \vdet\ is small. 
They could be either too young or too old to have an associated 
fast-expanding \schi shell. Their nature can be inferred in 
Figure~\ref{fig:snrs-rv2}, 
which is same as Figure~\ref{fig:snrs-rv}, but we now 
also mark the SNRs without detected \schi shells 
using their minimum velocities for detection. 
In Figure~\ref{fig:snrs-rv2},
an SNR without detected \schi can be either an old SNR with a velocity
less than \vdet\ or
a young SNR above the grid, i.e., with velocity greater than that
at $t/t_{\rm sf}=1$. Note that if an SNR is above the grid, it means that
the remnant is in adiabatic phase and does not have an \schi shell.
It is ``too young'', i.e., younger than $t_{\rm sf}$ in equation (6).
Note that $t_{\rm sf}$ is large when the ambient density is low.
For small SNRs, e.g., those with $R_s\le 10 - 20$~pc,
the latter possibility is more likely unless they are in a
dense environment, such as a molecular cloud.
For larger SNRs, both possibilities are likely.
Further observational studies will be helpful to address the nature of such
SNRs individually.

As a final comment, 
our study in this paper has targeted {\em known} SNRs,
but there could be many missing SNRs.
The estimated SN rate in our Galaxy ranges 
from 1.4 to 5.8 $\times 10^{-2}$~yr$^{-1}$ \citep{li11}.  
If we adopt the recent estimate from the Lick 
Observatory Supernova Search, 
$2.84 \pm 0.60\times 10^{-2}$~yr$^{-1}$ \citep{li11}, 
the total number of radio SNRs in the Galaxy would be 
$\sim 2800$, assuming $1\times 10^5$~yrs of visible radio continuum emission.
Then, simply using the geometrical fraction of the survey area, which is 18\%
for a Galactic disk radius of 15~kpc, 
the expected number of radio SNRs in the survey area is $\sim 500$.
Therefore, the number of known SNRs (39) is only 8\% of the expected population.
This small fraction could be due to several factors: 
faintness of old SNRs in radio continuum, 
confusion due to Galactic background emission,  
collective explosions of CCSNe that produce supershells
instead of SNRs, etc.\ \citep[see][]{koo04a, higdon05, brogan06}. 
So in principle, there could be many old SNRs or supershells not 
visible in radio continuum but visible in the \schi 21-cm line,
and it may be worthwhile to search for such \schi features. 

\section{Summary}
\label{sec:sum}

The \igalfa\ survey provides fully-sampled \schi data 
covering the Galactic plane between
longitudes 32 to 77 degrees and latitudes $-10$ to $+10$ degrees.
The high resolution (4\arcmin) and high sensitivity (0.2 K) of the 
data provide an opportunity to investigate small-scale, 
faint \schi emission in the diffuse ISM.
In this paper, we have explored the \igalfa\ 
data toward the all known 39 SNRs in order to search for 
associated fast-expanding \schi shells.
Our main results are as follows:

\noindent
1. Among the 39 SNRs in the survey area, four SNRs show  
associated high-velocity (HV) \schi emission.
These four SNRs were classified by KH91 as rank~3 SNRs with 
excess emission at the highest positive velocities 
in their low-resolution (30\arcmin) \schi study.
KH91 listed another SNR (W50) as a candidate with associated 
HV \schi emission out of 26 SNRs known in the survey area at that time. 
But the high-resolution \igalfa\ 
data show that the emission extends well beyond the  
SNR boundary, so we consider it not associated with the SNR. 
Surprisingly, we have not detected associated HV \schi emission 
in any of the ten SNRs discovered since the work of KH91.

\noindent
2. The four SNRs where we have detected physically associated HV \schi gas are 
G$34.7-0.4$ (W44), G$49.2-0.7$ (W51C), \g54 (HC40), and G$69.0+2.7$ (CTB~80).
Their velocity structures indicate that the 
HV \schi gas is in portions of expanding shells. 
In the SNR W44, we see \schi emission from 
both receding and approaching portions of the shell, which is  
the first ever such detection.
In the other SNRs, we could see only the receding portions of the shells. 
The SNR \g54 shows highly circularly symmetric \schi emission that 
matches very well with its radio continuum morphology. 
There is a ring structure lying just outside the SNR boundary, 
which could be a pre-existing structure formed by the progenitor. 
We discuss the properties of the expanding
\schi shells in these two SNRs and derive their physical parameters.
The other two SNRs have been studied previously in detail. 
The \igalfa\ results are consistent with those previous studies. 

\noindent
3. The four SNRs with associated fast-expanding \schi shells 
are all middle-aged SNRs with $t_s = 1.8 - 9.5$\ee4~yr (Table~\ref{tab:rank3}). 
The expansion velocities of the shells range from 59 to 135~\kms.
Notably,
their estimated ambient densities are all $\simgt 1$~cm$^{-3}$,
significantly
higher than that of the warm or hot ISM filling most of interstellar space. 
Two of them have associated PWN indicating that they are the remnants of CCSNe.
The other two do not have PWN but are interacting with molecular clouds,
so they are also likely the remnants of CCSNe. 
Therefore, the SNRs with \schi shells (\schi SNRs) that 
are detected are middle-aged SNRs of probable CCSN origin 
interacting with a relatively dense medium.
Large \schi SNRs in the diffuse ISM could be detected in principle, 
but they have not found.

\begin{figure}[ht]
\epsscale{0.7}
\plotone{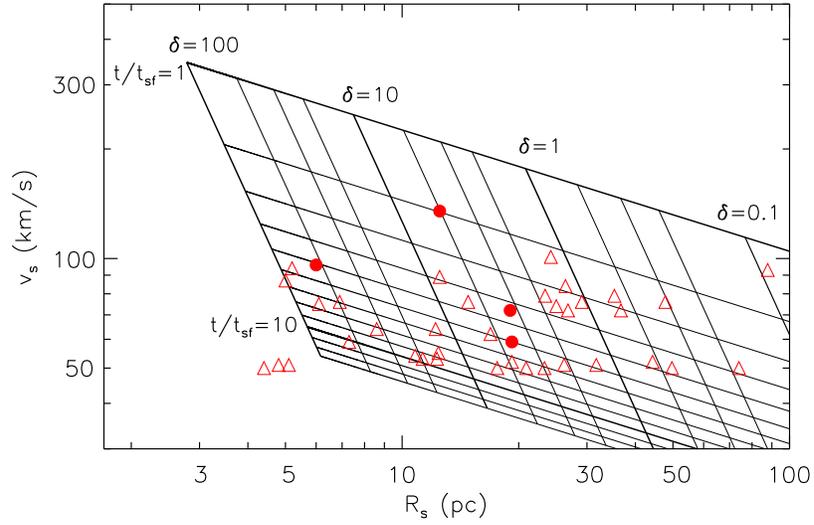}
\caption{
Same as Figure~\ref{fig:snrs-rv} but with all SNRs not seen in \igalfa\ \schi
also marked (empty triangles).
The velocities of the latter SNRs are the minimum velocities of detection for 
their hypothetical \schi shells.  The four SNRs with detected fast-expanding
\schi shells (filled circles) are shown with their measured expansion 
velocities. 
}
\label{fig:snrs-rv2}
\end{figure}

\noindent
4. The visibility of \schi SNRs depends on their location in the Galaxy. 
Three of the four detected \schi SNRs (excluding W44) are 
located where the visibility is favorable.
On the other hand, many 
SNRs are {\em not} detected in the \schi 21-cm line despite
having favorable locations.
They could be either too young or too old to have an associated 
fast-expanding \schi shell. We present a diagram (Figure~\ref{fig:snrs-rv2})
that can be used to infer the nature of these SNRs.

\acknowledgments

We thank the anonymous referee for the careful review and thoughtful comments on the paper.
The Inner Galaxy ALFA (I-GALFA\footnote{http://www.naic.edu/$^\sim$igalfa}) 
survey data are part of the Galactic ALFA (GALFA) \schi
project\footnote{https://sites.google.com/site/galfahi/} observed with the Arecibo L-band Feed Array (ALFA) on the 305-meter William E. Gordon Telescope.  The Arecibo Observatory is a U.S. National Science Foundation facility operated under sequential cooperative agreements with Cornell University and SRI International, the latter in alliance with the Ana G. M\'{e}ndez-Universidad Metropolitana and the Universities Space Research Association.
This work has been supported by the Korean Research
Foundation under grant KRF-2008-313-C00372 to B.-C. K.

{}


\begin{deluxetable}{lcccc cl}
\tabletypesize{\scriptsize}
\setlength{\tabcolsep}{0.03in}
\tablecaption{Supernova Remnants in the \igalfa\ Survey Area and Their Associated High-Velocity \schi Gases}
\tablewidth{0pt}
\tablehead{
\colhead{G-Name} & \colhead{Other Name} & \colhead{Size} & \colhead{Type} & \colhead{$T_{\rm b, 21cm}$} &
\colhead{KH91} & Note on the High-Velocity \schi Gas from This Work \\
\cline{6-6}
& & \colhead{(arcmin)} & & \colhead{(K)} & \colhead{Rank\tablenotemark{1}} &  
}
\startdata
G$31.9+0.0$& 3C391      & 7$\times$5    & S &      140~\phn\phn& 1      &  \nodata          \\
G$32.1-0.9$& \nodata    &~40?           &~C?&     \nodata      & \nodata&  \nodata          \\
G$32.4+0.1$& \nodata    & 6             & S &\phn\phn1.7?      & \nodata&  \nodata          \\
G$32.8-0.1$& Kes 78     & 17            &~S?&\phn\phn8.6?      & 2      &  \nodata          \\
G$33.2-0.6$& \nodata    & 18            & S &    \phn2.2       & 0      &  \nodata          \\
G$33.6+0.1$& Kes 79     & 10            & S &       45~~       & 2      &  \nodata          \\
G$34.7-0.4$& W44        & 35$\times$27  & C &       52~~       & 3      &  associated HV \schi gas at $+124$ -- $+240$ and $-120$ -- $-67$~\kms\ \\
G$35.6-0.4$& \nodata    & 15$\times$11  & S &    \phn9.7       & \nodata&  \nodata          \\
G$36.6-0.7$& \nodata    &~25?           &~S?&     \nodata      & 0      &  \nodata          \\
G$36.6+2.6$& \nodata    &~17$\times$13? & S &\phn\phn0.6?      & \nodata&  \nodata          \\
G$39.2-0.3$& 3C396      & 8x6           & C &       74~~       & 2      &  \nodata          \\
G$39.7-2.0$& W50        & 120$\times$60 & ? &\phn\phn2.2?      & 3      &  HV \schi gas at $+99$ -- $+124$~\kms, but extends beyond the SNR \\
G$40.5-0.5$& \nodata    & 22            & S &    \phn4.6       & (3)    &  HV \schi gas at $-124$ -- $-70$~\kms, but portion of a larger structure \\
G$41.1-0.3$& 3C397      & 4.5$\times$2.5& S &      401~\phn\phn& (3)    &  HV \schi gas at $+110$ -- $+120$~\kms, but probably background emission \\
G$42.8+0.6$& \nodata    & 24            & S &\phn\phn1.1?      & (3)    &  HV \schi gas at $+100$ -- $+108$~\kms, but probably background emission \\
G$43.3-0.2$& W49B       & 4$\times$3    & S &      649~\phn\phn& 1      &  \nodata          \\ 
G$43.9+1.6$& \nodata    &~60?           &~S?&\phn\phn0.5?      & \nodata&  \nodata          \\
G$45.7-0.4$& \nodata    & 22            & S &\phn\phn1.8?      & 0      &  \nodata          \\
G$46.8-0.3$& (HC30)     & 17$\times$13  & S &       13~~       & 1      &  \nodata          \\
G$49.2-0.7$& W51C       & 30            &~S?&       39?~       & 3      &  associated HV \schi gas at $+91$ -- $+160$~\kms\  \\
G$53.6-2.2$& 3C400.2    & 33x28         & S &    \phn1.6       & 0      &  \nodata          \\
G$54.1+0.3$\tablenotemark{2}                                       
           & \nodata    & 12.4          & C &   $1.9-3.8$      & 1      &  \nodata          \\
G$54.4-0.3$& (HC40)     & 40            & S &    \phn3.6       & 3      &  associated HV \schi gas at $+80$ -- $+130$~\kms\  \\
G$55.0+0.3$& \nodata    &~20$\times$15? & S &\phn\phn0.3?      & \nodata&  \nodata          \\
G$55.7+3.4$& \nodata    & 23            & S &    \phn0.5       & 0      &  \nodata          \\
G$57.2+0.8$& (4C21.53)  &~12?           &~S?&\phn\phn2.5?      & 0      &  \nodata          \\
G$59.5+0.1$& \nodata    & 15            & S &\phn\phn2.7?      & \nodata&  \nodata          \\
G$59.8+1.2$& \nodata    &~20$\times$16? & ? &    \phn1.0       & \nodata&  \nodata          \\
G$63.7+1.1$& \nodata    & 8             & F &    \phn6.1       & \nodata&  \nodata          \\
G$65.1+0.6$& \nodata    & 90$\times$50  & S &    \phn0.2       & \nodata&  \nodata          \\ 
G$65.3+5.7$& \nodata    & 310$\times$240&~S?&\phn\phn0.1?      & (3)    &  HV \schi gas at $-122$ -- $-159$~\kms, but no morphological relation \\
G$65.7+1.2$& DA 495     & 22            & F &    \phn2.1       & 2      &  \nodata          \\
G$67.7+1.8$& \nodata    & 15$\times$12  & S &    \phn1.1       & \nodata&  \nodata          \\
G$68.6-1.2$& \nodata    & 23            & ? &\phn\phn0.3?      & \nodata&  \nodata          \\
G$69.0+2.7$& CTB 80     &~80?           & ? &\phn\phn3.8?      & 3      &  associated HV \schi gas at $+43$ -- $+110$~\kms\   \\
G$69.7+1.0$& \nodata    & 16$\times$14  & S &    \phn1.7       & \nodata&  \nodata          \\
G$73.9+0.9$& \nodata    & 27            &~S?&    \phn2.7       & 0      &  \nodata          \\
G$74.0-8.5$& Cygnus Loop& 230$\times$160& S &    \phn1.2       & 0      &  \nodata          \\
G$74.9+1.2$& CTB 87     & 8$\times$6    & F &       38~~       & 0      &  \nodata          \\
\enddata
\tablecomments{
(1) The size is the angular diameter in radio continuum; 
    a single value is quoted for nearly circular remnants, 
    and the product of two values, the major and minor axes, is quoted for elongated remnants. 
(2) The type codes `S', `F', or `C' represent SNRs with a `shell', `filled-centre', or
    `composite' radio structure. Uncertain parameters are listed with a question mark.
(3) The mean brightness temperature at 21~cm $T_{\rm b, 21 cm}$ is 
    calculated from the 1~GHz flux ($F_{\rm 1 GHz}$) and spectral index $\alpha$ 
    ($F_\nu\propto \nu^\alpha$) in Green's catalog, i.e., 
    $T_b = 1.42^\alpha F_{\rm 1 GHz} \lambda^2 /(2k \Delta\Omega_S)$ 
    where $\lambda$ is 21.1~cm and $\Delta \Omega_S$ 
    is the solid angle of the source in steradians. 
    For sources without spectral indexes, we adopt $-0.5$.
    We use the area of a circle (or ellipse) 
    corresponding to the sizes in Table~\ref{tab:all} as the solid angle.
}
\tablenotetext{1}{Rank `(3)' SNRs were classified as `3'
      but regarded to be not-associated with the SNRs by KH91.}
\tablenotetext{2}{This SNR is classified as type `C' since a larger radio shell at 1.4~GHz is detected by 
      \citet{lang10}. Its size and surface brightness, which is 100-200 mJy beam$^{-1}$ with a beam of
      $6\farcs82\times6\farcs60$, are used to calculate $T_b$.} 
\label{tab:all}
\end{deluxetable}
\begin{deluxetable}{lcccc ccccc ccc}
\tabletypesize{\scriptsize}
\setlength{\tabcolsep}{0.03in}
\tablecaption{Physical Parameters of Supernova Remnants with Fast-Expanding \schi Shells}
\tablewidth{0pt}
\tablehead{
\colhead{Name} & \colhead{$d$} & \colhead{$R_{\rm s}$} & \colhead{$v_0$} & \colhead{$\vexp$} &
\colhead{Age} & \colhead{\schi Mass} & \colhead{K.E.} & \colhead{$n_{\rm 0}$} & \colhead{$\esn$} &
\colhead{P?\tablenotemark{1}} & \colhead{M?\tablenotemark{2}} \\
 & \colhead{(kpc)} & \colhead{(pc)} & \colhead{(\kms)} & \colhead{(\kms)} & \colhead{(\ee{4}~yr)} &
\colhead{(\msol)} & \colhead{(\ee{49}~erg)} & \colhead{(cm$^{-3}$)} & \colhead{(\ee{50}~erg)} &  & 
}
\startdata
W44  &2.8  &12.5  &47&135(2)&2.73(0.04)& 393(13) & 9.9(0.4)&1.9(0.1)  &3.2(0.1) &y      &y       \\
W51C\tablenotemark{3}  
     &6\phn& 6\phn&62& 96(6)&1.8(0.1)  &$>$1200  &$>$17    &$\sim 100$&$\sim 19$&\nodata&y       \\
\g54 &3.3  &19.2  &40& 59(6)&9.5(1.0)  & 580(150)& 2.8(0.9)&0.79(0.20)&1.5(0.5) &\nodata&y       \\
CTB~80\tablenotemark{4}
     &2\phn&18.6  &13& 72(3)&7.7(0.3)  &1050(210)& 7.6(1.5)&1.5(0.3)  &3.8(0.9) &y      &\nodata \\
\enddata
\tablecomments{
The distance ($d$), radius ($R_{\rm s}$), and systemic velocity ($v_0$) 
are adopted values, so that they do not have errors. 
The errors in W44 and \g54 are formal errors from the fit. 
}
\tablenotetext{1}{Detection (y) of associated pulsars. 
References are \citet{wolszczan91} and \citet{kulkarni88} for W44 and CTB~80, respectively.}
\tablenotetext{2}{Detection (y) of associated molecular clouds. 
References are \citet{wootten77}, \citet{koo97b}, and \citet{junkes92a}
for W44, W51C, and \g54, respectively.}
\tablenotetext{3}{The \schi shell parameters are from \citet{koo97a}. 
W51C is interacting with a molecular cloud, and
only a lower limit to the \schi mass has been obtained. 
We adopt 100~cm$^{-3}$ as a characteristic density of the cloud.  
}
\tablenotetext{4}{The \schi shell parameters are from \citet{koo90}. 
\cite{koo90} obtained an \schi mass of 1200 \msol\
by fitting a Gaussian profile to the observed
\schi mass distribution. But an independent estimate
of the mass of the shell is available from infrared studies
(900 \msol; see Koo et al. 1990). We adopt
the mean of the two as the mass of the expanding shell.
}
\label{tab:rank3}
\end{deluxetable}
\begin{deluxetable}{lccrc ccccc ccl}
\tabletypesize{\scriptsize}
\setlength{\tabcolsep}{0.03in}
\tablecaption{Distances to the SNRs in the \igalfa\ Survey Area}
\tablewidth{0pt}
\tablehead{
 && \multicolumn{2}{c}{Coordinates} && && \multicolumn{2}{c}{Distance} && & &  \\
\cline{3-4} \cline{8-9}
\colhead{G-Name}&& \colhead{$\ell$}& \colhead{\itb}&& \colhead{Type}&& 
\colhead{Quoted}& \colhead{$\Sigma-D$}&& \colhead{Radius}& \colhead{\vdet\tablenotemark{1}} &
\multicolumn{1}{l}{Reference(s)}\\
 && \colhead{(\arcdeg)}& \colhead{(\arcdeg)}&& && \colhead{(kpc)} & \colhead{(kpc)} && \colhead{(pc)} &
\colhead{(\kms)} &  
}
\startdata
G$31.9+0.0$&&31.89&   0.03&&     S &&       8.5                   &\phn5.4&&\phn7.3& 59&\citet{green09a}             \\
G$32.1-0.9$&&32.12&$-$0.90&&\phn C?&&       4.6                   &\nodata&&   26.8& 72&\citet{folgheraiter97}       \\
G$32.4+0.1$&&32.41&   0.11&&     S &&      17\phn\phn             &   33.9&&   14.8& 76&\citet{yamaguchi04}          \\
G$32.8-0.1$&&32.81&$-$0.06&&\phn S?&& $\sim$7.1\phn               &\phn6.3&&   17.6& 50&\citet{koralesky98,boumis09} \\
G$33.2-0.6$&&33.18&$-$0.55&&     S &&     \nodata\phn             &   10.1&&   26.4& 84&\nodata                      \\
G$33.6+0.1$&&33.70&   0.01&&     S &&       7.8\tablenotemark{2}  &\phn5.1&&   11.3& 53&\citet{frail89,green04,green09a} \\
G$34.7-0.4$&&34.67&$-$0.39&&     C &&       2.8\tablenotemark{2}  &\nodata&&   12.5& 89&\citet{caswell75,green04,green09a} \\
G$35.6-0.4$&&35.59&$-$0.50&&     S &&       3.7                   &\phn7.6&&\phn6.9& 76&\citet{green09b}             \\
G$36.6-0.7$&&36.59&$-$0.69&&\phn S?&&     \nodata\phn             &\nodata&&\nodata& 53&\nodata                      \\
G$36.6+2.6$&&36.58&   2.60&&     S &&     \nodata\phn             &   20.5&&   44.3& 52&\nodata                      \\
G$39.2-0.3$&&39.24&$-$0.32&&     C &&       8.5                   &\nodata&&\phn8.6& 64&\citet{lee09}                \\
G$39.7-2.0$&&39.69&$-$2.39&&     ? &&       6.0                   &\nodata&&   74.0& 50&\citet{green09a}             \\
G$40.5-0.5$&&40.52&$-$0.51&&     S &&       3.8\tablenotemark{3}  &\phn6.1&&   12.2& 64&\citet{yang06}               \\
G$41.1-0.3$&&41.11&$-$0.31&&     S &&      10.6\tablenotemark{3}  &\phn6.1&&\phn5.2& 94&\citet{jiang10}              \\
G$42.8+0.6$&&42.82&   0.64&&     S && $\sim$6\phn\phn\phn         &   10.3&&   20.9& 50&\citet{marsden01}            \\
G$43.3-0.2$&&43.27&$-$0.19&&     S &&      10\phn\phn             &\phn4.8&&\phn5.0& 87&\citet{green09a}             \\
G$43.9+1.6$&&43.91&   1.61&&\phn S?&&     \nodata\phn             &\phn5.7&&   49.7& 50&\nodata                      \\
G$45.7-0.4$&&45.69&$-$0.39&&     S &&     \nodata\phn             &\phn9.1&&   29.1& 76&\nodata                      \\
G$46.8-0.3$&&46.77&$-$0.30&&     S && $\sim$7.8                   &\phn5.8&&   16.9& 62&\citet{green09a}             \\
G$49.2-0.7$&&49.14&$-$0.60&&\phn S?&&       6\phn                 &\phn1.9&&   26.2& 51&\citet{koo95b}               \\
G$53.6-2.2$&&53.63&$-$2.26&&     S &&       2.8                   &\phn6.6&&   12.4& 55&\citet{giacani98,green09a}   \\
G$54.1+0.3$&&54.09&   0.26&&     C &&       6\phn                 &\nodata&&   10.8& 54&\citet{hjk13}                \\
G$54.4-0.3$&&54.47&$-$0.29&&     S &&       3.3\tablenotemark{4}  &\phn3.7&&   19.2& 52&\citet{junkes92a,case98}     \\
G$55.0+0.3$&&55.11&   0.42&&     S &&      14\phn\phn             &   23.1&&   35.3& 79&\citet{matthews98,green09a}  \\
G$55.7+3.4$&&55.60&   3.51&&     S &&     \nodata\phn             &   14.3&&   47.8& 76&\nodata                      \\
G$57.2+0.8$&&57.30&   0.83&&\phn S?&&     \nodata\phn             &   14.3&&   25.0& 74&\nodata                      \\
G$59.5+0.1$&&59.58&   0.12&&     S &&     \nodata\phn             &   11.1&&   24.2&101\phn&\nodata                      \\
G$59.8+1.2$&&59.81&   1.20&&     ? &&     \nodata\phn             &   14.1&&   36.7& 72&\nodata                      \\
G$63.7+1.1$&&63.79&   1.17&&     F &&     \nodata\phn             &\nodata&&\nodata&\nodata&\nodata                      \\
G$65.1+0.6$&&65.27&   0.30&&     S &&       9\phn                 &\phn6.8&&   87.8& 93&\citet{green09a}             \\
G$65.3+5.7$&&65.18&   5.66&&\phn S?&&       0.8                   &\phn2.1&&   31.7& 51&\citet{boumis04,green09a}    \\
G$65.7+1.2$&&65.72&   1.21&&     F &&       1.5                   &\nodata&&\phn4.8& 51&\citet{kothes04,green09a}    \\
G$67.7+1.8$&&67.74&   1.82&&     S &&$\sim$12\phn\phn\phn\phn     &   18.0&&   23.4& 79&\citet{mavromatakis01}       \\
G$68.6-1.2$&&68.60&$-$1.20&&     ? &&     \nodata\phn             &   19.2&&   64.2& 50\tablenotemark{5}&\nodata                      \\
G$69.0+2.7$&&68.84&   2.78&&     ? &&       2\phn                 &\phn1.8&&   23.3& 50&\citet{koo90}                \\
G$69.7+1.0$&&69.69&   1.00&&     S &&   \phn2\tablenotemark{6}\phn&   13.2&&\phn4.4& 50&\citet{yoshita00}            \\
G$73.9+0.9$&&73.91&   0.88&&\phn S?&& $\sim$1.3\phn               &\phn6.4&&\phn5.1& 51&\citet{lozinskaya93}         \\
G$74.0-8.5$&&73.98&$-$8.56&&     S &&   \phn0.44                  &\phn1.2&&   12.3& 53&\citet{green09a}             \\
G$74.9+1.2$&&74.94&   1.14&&     F &&       6.1                   &\nodata&&\phn6.1& 75&\citet{kothes03,green09a}    \\
\enddata
\tablecomments{
 Distance with a symbol of `$\sim$' denotes that 
 we adopt the average of possible distances given by reference(s). 
}
\tablenotetext{1}{Minimum expansion velocity for detection.}
\tablenotetext{2}{Recalculated by \citet{green04} assuming a flat rotation curve
                  with $R_\odot = 8.5$~kpc and $\Theta_\odot = 220$~\kms.}
\tablenotetext{3}{Recalculated by this work using the Galactic rotation curve of \citet{brand93} 
                  with $R_\odot = 8.5$~kpc and $\Theta_\odot = 220$~\kms.}
\tablenotetext{4}{Recalculated by \citet{case98} assuming a flat rotation curve
                  with $R_\odot = 8.5$~kpc and $\Theta_\odot = 220$~\kms.}
\tablenotetext{5}{This remnant is outside the assumed disk radius of 15~kpc,
                  so we simply adopt 50~\kms\ as the minimum velocity.}
\tablenotetext{6}{\citet{yoshita00} suggested that G$69.7+1.0$ will be at a similar distance as CTB~80,  
                  since the column density of the ISM between us and G$69.7+1.0$ is analogous with that of CTB~80.
                  In this paper, we adopt 2~kpc for both CTB~80 and G$69.7+1.0$.}
\label{tab:distance}
\end{deluxetable}
%

%

%

%

%
%
%

\end{document}